\documentclass[pdflatex,sn-mathphys-num]{sn-jnl}

\usepackage{graphicx}%
\usepackage{multirow}%
\usepackage{amsmath,amssymb,amsfonts}%
\usepackage{amsthm}%
\usepackage{mathrsfs}%
\usepackage[title]{appendix}%
\usepackage{xcolor}%
\usepackage{textcomp}%
\usepackage{manyfoot}%
\usepackage{booktabs}%
\usepackage{algorithm}%
\usepackage{algorithmicx}%
\usepackage{algpseudocode}%
\usepackage{listings}%
\usepackage{cleveref}
\usepackage{a4wide}
\usepackage{orcidlink}
\usepackage[exponent-product = \cdot]{siunitx}
\usepackage{subcaption}
\usepackage[normalem]{ulem}

\newif\ifrevisioncolor
\revisioncolorfalse   
\ifrevisioncolor
  \newcommand{\revision}[1]{\textcolor{blue}{#1}}
\else
  \newcommand{\revision}[1]{#1}
\fi

\newif\ifrevisioncolorTwo
\revisioncolorTwofalse   
\ifrevisioncolorTwo
  \newcommand{\revisionTwo}[2]{\textcolor{red}{\sout{#1}}\,\textcolor{blue}{#2}}
\else
  \newcommand{\revisionTwo}[2]{#2}
\fi

\theoremstyle{thmstyleone}%
\theoremstyle{thmstyletwo}%

\theoremstyle{thmstylethree}%

\begin{document}

\title{Augmenting a pure and hybrid vertical equilibrium scheme via data-driven surrogate modelling}


\author*[1]{\fnm{Ivan} \sur{Buntic}\orcidlink{0000-0001-9376-1260}}\email{ivan.buntic@iws.uni-stuttgart.de}

\author[1]{\fnm{Bernd} \sur{Flemisch}\orcidlink{0000-0001-8188-620X}}\email{bernd.flemisch@iws.uni-stuttgart.de}

\affil[1]{\orgdiv{Department of Hydromechanics and Modelling of Hydrosystems}, \orgname{University of Stuttgart}}


\abstract{Vertical equilibrium (VE) models have been introduced as computationally efficient alternatives to traditional mass and momentum balance equations for fluid flow in porous media. Since VE models are only accurate in regions where phase equilibrium holds\revisionTwo{, while}{ and} traditional simulations are computationally demanding, hybrid methods have been proposed to combine the accuracy of the full-dimensional approach with the efficiency of VE model. However, coupling both models introduces computational overhead that can make hybrid simulations slower than fully traditional ones. \revisionTwo{To address this,}{To address the computational overhead introduced by coupling interfaces in hybrid models,} we \revisionTwo{introduce}{utilize} data-driven surrogates to \revisionTwo{}{accelerate the overall scheme. To this end, we} predict the gas plume distance and coarse-level mobilities in the VE model\revisionTwo{}{, and also enhance the computation of the coupling scheme via surrogates}. We focus on surrogate models with short inference times to minimize computational overhead during frequent function calls. The proposed approach preserves key physical properties, such as mass conservation\revisionTwo{}{, despite the deployment of data-driven models,} while substantially reducing simulation runtimes. Overall, combining data-driven methods with the hybrid VE scheme yields an enhanced model that outperforms traditional simulations in speed while introducing only negligible errors.}




\maketitle

\noindent\textbf{Article Highlights}
\begin{itemize}
  \item Optimizing individual, algorithmic components of a hybrid VE scheme via data-driven surrogates.
  \item Computational speedup of \revision{up to 75\%}\revisionTwo{ while preserving physical properties such as mass conservation.}{.}
  \item \revisionTwo{Open-source implementation in C++}{Preservation of physical properties such as mass conservation}.
\end{itemize}

\section{Introduction}
The simulation of underground storage sites for gaseous and non-gaseous fluids has become an increasingly prominent topic in recent years~\cite{gieeu2021,alshafi2023}. Established approaches, such as carbon capture and storage (CCS), as well as more recent strategies that utilize subsurface reservoirs as dynamic storage sites for gaseous energy carriers, \revision{such as methane or hydrogen,} benefit greatly from reliable and representative simulation tools. \revisionTwo{In practice, three main types of formations are considered as potential storage sites. Salt caverns enable the controlled storage of gases~\cite{liu2023}, although their total capacity is limited. Depleted oil and gas fields~\cite{wan2024} offer larger volumes and the advantage of well-documented reservoir properties due to prior extraction activities. Moreover, depleted gas fields typically feature an impermeable caprock~\cite{aluah2024}, which prevents buoyant fluids from escaping. Despite these advantages, their total potential capacity is estimated to be only {1–5\%}~\cite{luo2022} of the potential storage capacity of underground saline aquifers~\cite{celia2015}, which represent the third option. Saline aquifers, while offering the largest storage capacities, are more difficult to identify as suitable storage sites since they must meet strict geological criteria to ensure containment. This underscores the importance of reliable simulation frameworks to provide insight into the processes occurring during storage.}{}Traditional simulation techniques~\cite{malay2018}, based on the full solution of mass and momentum balance equations, have long been employed for this purpose. However, the computational cost of such models can be prohibitive. To address this, a variety of reduced-order models have been proposed, aiming to balance efficiency and accuracy. Among these, the vertical equilibrium (VE) method~\cite{gasda2011,nordbotten2011} has attracted considerable attention. \revisionTwo{}{The vertical equilibrium model assumes instantaneous phase segregation in the vertical direction and is therefore only an appropriate approximation for systems dominated by buoyancy forces\cite{vahabzadeh2025}. To satisfy this assumption, the fluid phases must exhibit sufficiently large density contrasts, such that vertical segregation occurs on a time scale that is significantly shorter than that of lateral phase transport. }By integrating the \revisionTwo{}{traditional, }governing equations over the vertical dimension, the VE method reduces the spatial dimensionality of the problem, leading to a significant speedup. However, this simplification comes at the cost of reduced applicability since the method yields reliable results only in regions where vertical dynamic effects are negligible and the multiphase system is in equilibrium. \newline
\revisionTwo{}{The vertical equilibrium method is widely used for the simulation of long-term carbon dioxide sequestration~\cite{celia2015,gasda2009,gasda2011,court2012}, as the time scale of storage is significantly larger than that of phase segregation. \revisionTwo{}{Nevertheless, recent studies have shown that the VE method can, under certain conditions, also provide sufficiently accurate results for simulations of dynamic gas storage, such as hydrogen storage~\cite{vahabzadeh2025}.} Due to its comparatively low computational cost, the VE model is also employed for sensitivity analysis and parameter estimation~\cite{alamara2025,nilsen2017,zeynolabedini2025}, where repeated model evaluations are required.} \newline
In order to combine the efficiency of the VE method with the accuracy of full-dimensional formulations, hybrid modelling approaches have been developed~\cite{becker2022,moyner2019}. These approaches \revisionTwo{}{utilize multiple models that can differ in the physics they represent or the scale they operate on. Since several models are deployed at once, hybrid models critically rely on the coupling strategy between the respective submodels in order to exchange information across coupling interfaces. However}, the coupling procedure and the two-level formulation inherent in VE methods introduce a computational overhead. In some cases, this overhead may offset the expected benefits and render the hybrid simulation even more expensive than traditional full-dimensional approaches~\cite{buntic2025}.

In recent years, data-driven methods have emerged as powerful tools for constructing surrogate models that can replace or augment the direct computation of governing equations and physical relations~\cite{sun2020,kavana2024,chen2024,esfandi2023}. At the same time, their application requires careful consideration, since the predictive quality of such models is inherently limited by the quality and representativeness of the training data~\cite{wen2022}. If employed indiscriminately, machine learning approaches may yield unphysical predictions and misleading results. In this work, we employ data-driven models to enhance both a pure VE model and a hybrid formulation, with the objective of achieving consistently faster simulations without compromising accuracy or physical constraints. \revisionTwo{}{To this end, we first identify the algorithmic components that incur the highest computational cost and subsequently accelerate them using data-driven models. In particular, the coupling fluxes in hybrid schemes and the computation of coarse-level constitutive relations in the VE scheme are selected as targets for surrogate modelling.} It should be emphasized that no new machine learning methodology is proposed in this work. Rather, the focus lies on demonstrating how existing data-driven approaches can \revision{optimize individual blocks of established physical models to improve computational performance \cite{schultzendorff2024}}.

In the following, we will first introduce the methodology in \cref{ch_methodology}. We present the full-dimensional model, the vertical equilibrium model, and the coupling scheme necessary for the hybrid model in \cref{ch_math_models}. Then, in \cref{ch_data_driven_models}, the linear regression model as well as a spline interpolation method are showcased as the data-driven approaches for possible surrogates. In \cref{ch_results}, we discuss advantages and disadvantages of applying data-driven models to enhance the computation of the gas plume distance in \cref{ch_predict_zp}, the coarse-level mobilities in \cref{ch_predict_coarse_mobilities} and the phase densities and viscosities in \cref{ch_predict_secondaries}. Subsequently, we analyse which effects can be observed by deploying all presented surrogate models at once in \cref{ch_combined_surrogates}, followed by an exhibition on the limitations of the enhanced models in \cref{ch_limitation}. In the end, we draw conclusions from the presented results while also pointing to possible future extensions in \cref{ch_conclusion}. \revision{Finally, in the appendix \cref{ch_closed_form_zp} we present alternative, closed solutions to the gas plume distance before motivating the usage of linear regression and spline interpolation models in \cref{ch_inferencetime}.}

\section{Methods} \label{ch_methodology}
In this section, we present the mathematical frameworks for modelling fluid flow in porous media alongside the data-driven techniques employed as surrogate models. We begin by outlining the two-phase, compressible, immiscible Darcy equations, which define the full-dimensional (FD) model, and subsequently describe their adaptation to the vertical equilibrium (VE) formulation. We then address the coupling scheme~\cite{becker2018} that enables the construction of a hybrid approach through spatially dynamic model distribution. Finally, we introduce the data-driven components used to substitute selected algorithmic elements, focusing on linear regression and spline interpolation as representative surrogate models.

\subsection{Mathematical models} \label{ch_math_models}
In the following, we will briefly present the utilized, state-of-the-art flow equations which are based on mass and momentum conservation. A more detailed presentation of the models can be found in~\cite{buntic2025}.

\subsubsection{Full-dimensional model}
Our computations are based on an immiscible, compressible two-phase flow model which consists of mass balances
\begin{equation} \label{eq_fullD_mass}
    \frac{\partial \phi \varrho_\alpha s_\alpha}{\partial t} + \nabla \cdot \left(\varrho_\alpha \boldsymbol{u}_\alpha \right) = \varrho_\alpha q_\alpha, \quad \alpha\in \{w,n\},
\end{equation}
and momentum balances in the form of the Darcy equation
\begin{equation} \label{eq_fullD_momentum}
    \boldsymbol{u}_\alpha = -\boldsymbol{k} \lambda_\alpha \left( \nabla p_\alpha - \varrho_\alpha g_z \nabla z\right), \quad \alpha\in \{w,n\},
\end{equation}
for a wetting fluid phase $w$ and a non-wetting fluid phase $n$. Here, $\phi$ stands for the porosity of the porous medium, $\varrho_\alpha$ describes the phase density, $s_\alpha$ is the phase saturation, $\boldsymbol{u}_\alpha$ the Darcy velocity and $q_\alpha$ the source term for each phase. Additionally, $\boldsymbol{k}$ is the intrinsic permeability tensor, while $\lambda_\alpha$ describes the phase-specific mobility, $p_\alpha$ the phase pressure, $g_z$ the gravitational acceleration in $z$ direction and finally $z$ stands for position in vertical direction, positive pointing upward. Throughout all models, the wetting-phase pressure as well as the non-wetting-phase saturation are chosen to be the primary variables, while all \revisionTwo{other}{}other quantities can be deduced from these either via physical constraints or via empirical relations. As closing conditions, we utilize the Brooks-Corey capillary-pressure saturation relationship~\cite{brooksCorey1964},
\begin{equation} \label{eq_BrooksCorey_pc}
    p_n - p_w = p_c(s_w) = p_e s_{w,e}^{-1/\lambda} = p_e \left( \frac{s_w - s_{w,r}}{1-s_{w,r}-s_{n,r}} \right)^{-1/\lambda},
\end{equation}
with $p_c$ being the capillary pressure, $p_e$ the entry pressure, $s_{w,e}$ the effective saturation which is computed via the residual saturations $s_{w,r}$ and $s_{n,r}$. It is important to distinguish between the Brooks-Corey modelling parameter $\lambda$ and the phase mobility $\lambda_{\alpha}$. Additionally, we also require the sum of the saturations to equal one,
\begin{equation} \label{eq_sum_saturations}
    s_w + s_n = 1,
\end{equation}
so that the pore space is always filled by some fluid. From here on, we abbreviate the full-dimensional model with FD model. The computational domain $\Omega$ and the solution of the FD model are part of the space $\mathbb{R}^d \text{ with } d \in \lbrace 1,2,3\rbrace$. Additionally, for the computation of the phase mobilities, we first calculate the relative mobilities according to the Brooks-Corey law~\cite{brooksCorey1964}, 
\begin{align} 
    k_{rw}(s_w) &= \left(s_{w,e} \right)^{3 + 2/\lambda} \label{eq_relPermSat_w}, \\
    k_{rn}(s_w) &= \left(1-s_{w,e}\right)^2 \left(1-\left( s_{w,e} \right)^{1 + 2/\lambda} \right). \label{eq_relPermSat_n}
\end{align}
Finally, the relative permeability is divided by the dynamic viscosity $\mu_\alpha$ to obtain the phase mobility~\cite{zhang2002},
\begin{equation} \label{eq_phaseMobility}
    \lambda_\alpha = \frac{k_{r\alpha}}{\mu_\alpha}.
\end{equation}

\subsubsection{Vertical equilibrium model} \label{ch_VE_model}
\begin{figure}[hbt]
    \begin{center}
        \includegraphics[width=0.7\textwidth]{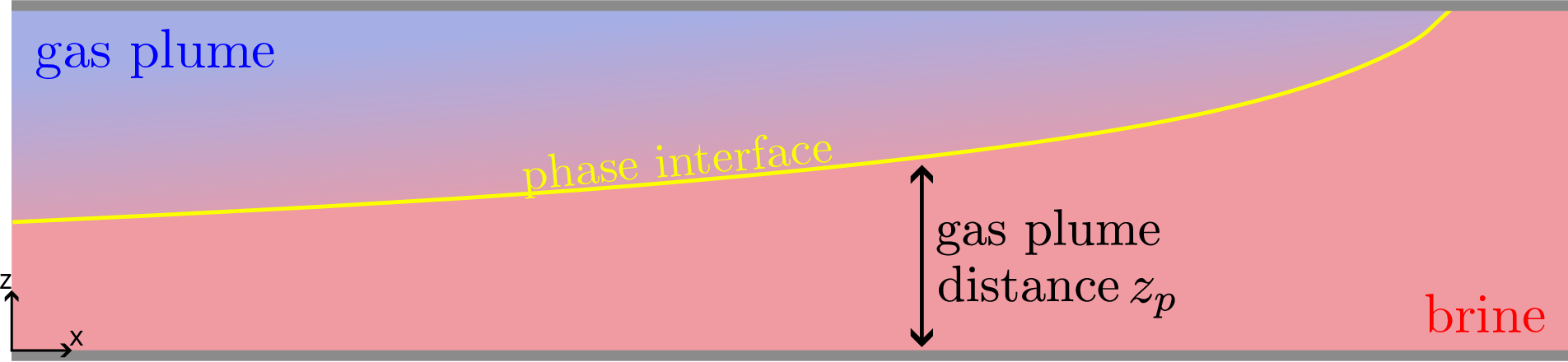}
    \end{center}
    \caption{Illustration of the gas plume distance $z_p$ in a 2D slice for a gas plume with caprocks at the top and bottom}
    \label{fig_gas_plume_dist}
\end{figure}

While the FD model operates on the space $\mathbb{R}^d$, the VE model operates on a space that is one dimension lower, that is $\mathbb{R}^{d-1}$. The VE equations~\cite{gasda2011} are obtained by integrating \cref{eq_fullD_mass} and \cref{eq_fullD_momentum} from the FD model over the vertical direction, thus reducing the effective spatial dimension by one. The integration leads to the mass balance equations
\begin{equation} \label{eq_VE_mass}
    \frac{\partial \varrho_\alpha \Phi S_\alpha}{\partial t} + \nabla_{//} \cdot \left(\varrho_\alpha \boldsymbol{U}_\alpha \right) = Q_\alpha,
\end{equation}
and the reduced form of the momentum equations
\begin{equation} \label{eq_VE_momentum}
    \boldsymbol{U}_\alpha = - \boldsymbol{K} \Lambda_\alpha \left( \nabla_{//} P_\alpha - \varrho_\alpha g_z \nabla_{//} z_B \right),
\end{equation}
which both structurally resemble \cref{eq_fullD_mass} and \cref{eq_fullD_momentum}. The capital letters denote a vertically integrated version of the quantities already discussed in the FD model,
\begin{align} \label{eq_integratedQuantities}
	\Phi &= \int_{z_B}^{z_T}{\phi \, \mathrm{d}z},      &      \boldsymbol{U}_\alpha &= \int_{z_B}^{z_T} {\boldsymbol{u}_\alpha \, \mathrm{d}z} , \notag \\
	S_\alpha &= \frac{1}{\Phi} \int_{z_B}^{z_T}{\phi s_\alpha \,\mathrm{d}z},     &      Q_\alpha &= \int_{z_B}^{z_T}{\varrho_\alpha q_\alpha \, \mathrm{d}z}, \notag \\
    \boldsymbol{K} &= \int_{z_B}^{z_T}{\boldsymbol{k}_{//} \,\mathrm{d}z},       &      \Lambda_\alpha &= \boldsymbol{K}^{-1} \int_{z_B}^{z_T}{\boldsymbol{k}_{//} \tilde{\lambda}_\alpha \, \mathrm{d}z},
\end{align}
while the subscript ${}_{//}$ describes operators that live on the reduced space. The top and bottom boundary of the computational domain are labelled as $z_T$ and $z_B$ respectively. \newline
Due to its simplifications, the VE model yields reliable results only in the absence of vertical dynamics. This condition corresponds to a complete segregation of the two phases, or equivalently, to a hydrostatic pressure distribution. Under such an assumption, it is possible to reconstruct a solution in $\mathbb{R}^d$, which we refer to as the fine level of the VE model, based on a solution in $\mathbb{R}^{d-1}$, denoted as the coarse level. A key element of this reconstruction~\cite{becker2021} is the computation of the so-called gas plume distance $z_p$, which defines the vertical position separating the two phases, as illustrated in \cref{fig_gas_plume_dist}.
The gas plume distance is obtained by balancing the phase masses of the fine-level and coarse-level solutions in the vertical direction:
\begin{align}\label{eq_gasPlumeDist_massBalance}
    \int_{z_B}^{z_T}{\tilde{s}_w(z) \, \mathrm{d}z} &= S_w (z_T - z_B).
\end{align}
Here, $\tilde{s}_w$ denotes the reconstructed saturation denoted by the tilde. By rearranging \cref{eq_BrooksCorey_pc} for the saturation $s_w$ and substituting it into \cref{eq_gasPlumeDist_massBalance} the resulting condition can be expressed as the non-linear balance
\begin{align} \label{eq_gasPlumeDist_integral_extended}
    z_p - z_B &+ \frac{1}{1-\lambda} \left( p_e + \left( \varrho_w - \varrho_n \right)g_z\left( z_T-z_p \right)\right)^{1-\lambda} \frac{\left( 1 - s_{w,r} - s_{n,r}\right) p_e^{\lambda}}{\left( \varrho_w - \varrho_n\right)g_z} \nonumber \\
    &- \frac{p_e \left( 1 - s_{w,r} - s_{n,r} \right)}{\left( 1-\lambda\right) \left( \varrho_w - \varrho_n \right) g_z } + s_{w,r}\left( z_T - z_p \right) - S_w (z_T - z_B) & = 0,
\end{align}
obtained by splitting the integral into two segments, from $z_B$ to $z_p$ and from $z_p$ to $z_T$.
The reconstructed saturation is defined by assuming a purely wetting phase from the bottom of the domain up to the gas plume distance, and a coexistence of the wetting and non-wetting phase between $z_p$ and the top of the domain. \revision{Under certain conditions, a closed solution for $z_p$ can be derived \cite{nordbotten2011b}. However, for arbitrary uniformity parameters $\lambda$, solving \cref{eq_gasPlumeDist_integral_extended} for $z_p$ is non-trivial and requires a numerical approximation. In \cref{ch_closed_form_zp}, we identify the values of $\lambda$ for which \cref{eq_gasPlumeDist_integral_extended} admits a closed-form solution.}

\subsubsection{Coupling scheme}
\begin{figure}[hbt]
    \begin{center}
        \includegraphics[width=0.4\textwidth]{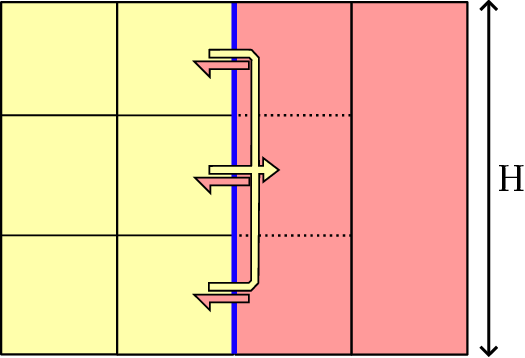}
    \end{center}
    \caption{\revision{Coupling concept between the FD (yellow) and the VE (red) domain across a coupling interface (blue line) with domain height $H=z_T-z_B$. The coupling fluxes are computed on the fine level and if necessary accumulated to a coarse-level flux entering the VE subdomain.}}
    \label{fig_coupling_graph}
\end{figure}

To construct a hybrid scheme that simultaneously employs the FD and VE model, appropriate coupling conditions at the interface between the two domains are required~\cite{becker2018}. These conditions are designed to ensure mass conservation by enforcing consistency between the FD quantities and the reconstructed VE quantities. Specifically, continuity of flux across the interface is imposed,
\begin{equation} \label{eq_coupling_flux}
    \boldsymbol{u}_{\alpha,c} = -k_c \lambda_{\alpha,c} \left( \nabla p_{\alpha,c} + \varrho_{\alpha,c}g\nabla\mathrm{z} \right),
\end{equation}
thereby preserving mass both locally and globally. The interface quantities, denoted with the subscript ${}_c$, are influenced by both domains. The permeability $k_c$ is evaluated using a harmonic average, the mobility \revisionTwo{$\lambda{\alpha,c}$}{$\lambda_{\alpha,c}$} is determined by an upwinding scheme, the pressure gradient incorporates contributions from both domains, and the density $\varrho_{\alpha,c}$ is obtained through arithmetic averaging. \revision{As depicted in \cref{fig_coupling_graph}, the coupling fluxes are computed on the fine level. Additionally, for the flux into a VE domain, the fine-level fluxes are accumulated into a single, coarse-level flux across the domain height $H=z_T-z_B$. A detailed explanation of the coupling strategy can be found in \cite{buntic2025}.}

\subsection{Data-driven methods} \label{ch_data_driven_models}
Data-driven models provide an attractive means of replacing computationally expensive algorithmic components with faster surrogates, thereby enabling significant computational speedups. A wide range of such methods exists, each with distinct advantages with their suitability depending on the specific requirements of the application. In the present work, where the goal is to accelerate both pure and hybrid VE schemes through data-driven schemes \revisionTwo{}{by only replacing individual, algorithmic components}, the primary criteria are high predictive accuracy and short inference times. Among available methods, linear regression and \revision{spline interpolation models} stand out for their exceptionally fast evaluations, \revision{which is demonstrated in \cref{ch_inferencetime}}.

\revision{Possible alternatives that are widely employed in reservoir engineering include operator-learning techniques \cite{wen2022b} and operator-based linearization methods \cite{voskov2017,khait2017}. Operator-learning approaches are well suited for capturing complex non-linear mappings and are particularly effective when predicting high-dimensional outputs, such as full solution fields \revisionTwo{of}{by replacing whole} partial differential equations~\cite{zhang2023,kazemi2023}. These methods can \revisionTwo{incorporate heterogeneous flow conditions and}{} achieve excellent accuracy at a very low evaluation cost. However, their use typically necessitates hundreds or even thousands of forward simulations to generate sufficiently rich training data that span the relevant input space. Since our objective is to predict scalar quantities rather than high-dimensional fields, such methods may not provide a meaningful advantage in our setting.}

\revision{Conversely, operator-based linearization techniques focus on local, linear approximations of non-linear operators within the numerical scheme. These approaches can be highly effective for intricate or strongly non-linear operators but offer limited benefits when applied to the comparatively simple relationships considered in this work. In cases where very high accuracy is required, we instead employ spline interpolation, which provides a local, bicubic approximation. In this work, we predominantly use linear regression models because of their low inference cost, and resort to spline interpolation only in cases where increased accuracy is essential.}

\subsubsection{Linear regression}
Linear regression is a classical machine learning technique for approximating observed data by linear combinations of polynomial functions of the input variables. The method is well documented in the literature~\cite{bishop2006,murphy2012}, but we briefly recall the formulation here to establish notation. We denote the input variables as features and the target quantity as $y$. In our setting, multiple features are used to predict a single target. The relation between features and the target can be written as a linear combination 
\begin{equation} \label{eq_linreg_matrix}
    \boldsymbol{y} = \boldsymbol{x} \boldsymbol{\beta} + \boldsymbol{\epsilon},
\end{equation}
\revision{in matrix formulation. Here, $\boldsymbol{y} \in \mathbb{R}^N$ represents the target vector, $\boldsymbol{x} \in \mathbb{R}^{N \times (d+1)}$ the feature matrix, $\boldsymbol{\beta} \in \mathbb{R}^{(d+1)}$ the coefficient vector, and $\boldsymbol{\epsilon}$ the error vector with $N\in\mathbb{N^+}$ being the number of observations and $d\in\mathbb{N^+}$ the number of features. Each row in \cref{eq_linreg_matrix} corresponds to one observation while the first entry $\beta^0$ yields the intercept of the linear regression.} Minimizing the error $\boldsymbol{\epsilon}$ in \cref{eq_linreg_matrix} yields the best-fit coefficients $\boldsymbol{\hat{\beta}}$ and the corresponding prediction
\begin{equation} \label{eq_linreg_prediction}
    \boldsymbol{\hat{y}} = \boldsymbol{x} \boldsymbol{\hat{\beta}},
\end{equation}
where $\boldsymbol{\hat{y}}$ denotes the approximated target. The optimal coefficients are determined by solving
\begin{equation} \label{eq_linreg_beta_hat}
    \boldsymbol{\hat{\beta}} = \underset{\boldsymbol{\beta}}{\arg\min} \, L(\boldsymbol{x}, \boldsymbol{y}, \boldsymbol{\beta}).
\end{equation}
While many estimation methods exist for minimizing \revision{the loss function $L$}, we employ least squares due to simplicity. 
One of the main reasons for selecting linear regression in this work is its computational efficiency. Predicting a target value requires only $d+1$ multiplications and $d$ additions per inference. Furthermore, the linear formulation guarantees a continuous dependence of the target on the input features, a property that is essential when approximating continuous quantities such as the gas plume distance (\cref{ch_predict_zp}) or \revision{secondary variables, such as density and viscosity. In general, secondary variables refer to quantities that can be computed from the primary solution variables (\cref{ch_predict_coarse_mobilities,ch_predict_secondaries})}.

\subsubsection{Spline interpolation}
In this section, we introduce bicubic spline interpolation, which maps a two-dimensional input space to a one-dimensional target space through bicubic functions \revision{via local interpolation}. The method is comprehensively documented in~\cite{deboor}, but is briefly summarized here to establish notation.

Consider a rectilinear grid with $N$ samples in the $x$-direction and $M$ samples in the $y$-direction, forming the vertices of a mesh. Each vertex is associated with an observed value $z_{ij}$, so that the mesh consists of points $(x_i,y_j, z_{ij})$, with $i\in{1,2,\dots,N}$ and $j\in{1,2,\dots,M}$. We refer to each element of this mesh as a patch. Within each patch, a local bicubic function $u_{ij}(x,y)$ is defined as
\begin{equation} \label{eq_general_bicubic}
    u_{ij}(x,y) = \sum_{m=0}^3 \sum_{n=0}^3 \alpha_{mn,ij} \left(x-x_{i-1} \right)^m \left(y - y_{j-1} \right)^n.
\end{equation}
This local function is valid on the patch $[x_{i-1},x_i]\times[y_{j-1}, y_j]$ and introduces 16 unknown coefficients $\alpha_{mn,ij}$ \revision{which can be solved for by evaluating the local interpolator and its partial derivatives at each local mesh vertex.}

Although spline interpolation is not typically categorized as a data-driven approach, it fits naturally into this framework. Constructing the global spline requires a dataset from which the coefficients of the local patches are determined. This procedure closely parallels the training stage of a data-driven model, where input features are selected and used to predict a target quantity. Once trained, predictions are generated by evaluating the linear combination of features, with the additional step of identifying the corresponding spline patch for the input. This extra step results in a slightly higher inference time compared to linear regression.

\section{Results} \label{ch_results}
\begin{figure}[h!]
    \centering
    \begin{subfigure}{\textwidth}
        \centering
        \includegraphics[width=\textwidth]{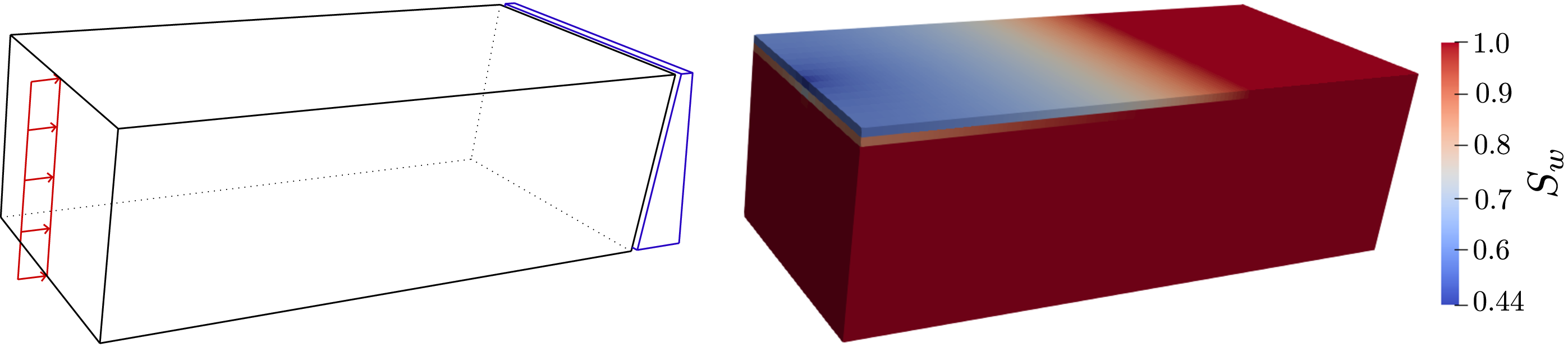}
        \caption{\revision{Pure VE injection test. No impermeable lenses. 130x16x24 grid elements. Domain height $H=\qty{80}{\meter}$}}
        \label{fig_case_a}
    \end{subfigure}

    \vspace{1em}

    \begin{subfigure}{\textwidth}
        \centering
        \includegraphics[width=\textwidth]{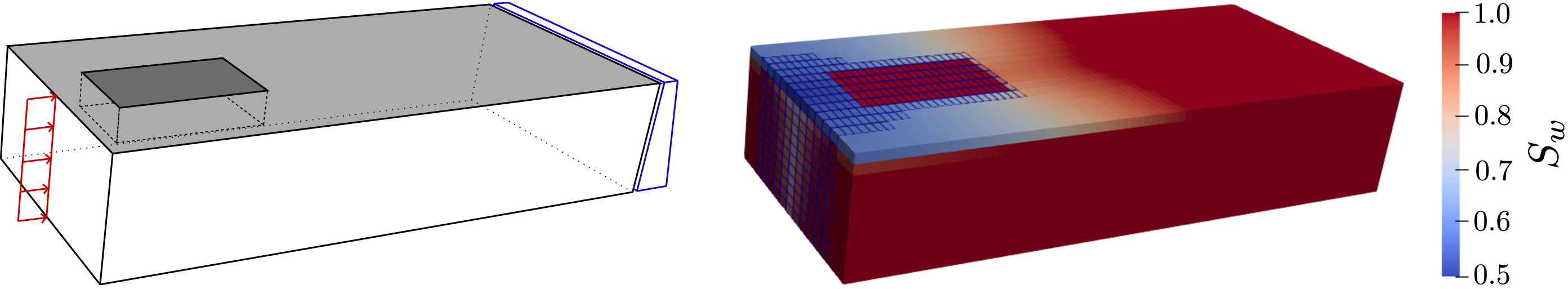}
        \caption{\revision{Hybrid, statically-coupled injection test. One impermeable lens (dark grey) at the top. 80x15x12 grid elements. Domain height $H=\qty{50}{\meter}$}}
        \label{fig_case_b}
    \end{subfigure}

    \vspace{1em}

    \begin{subfigure}{\textwidth}
        \centering
        \includegraphics[width=\textwidth]{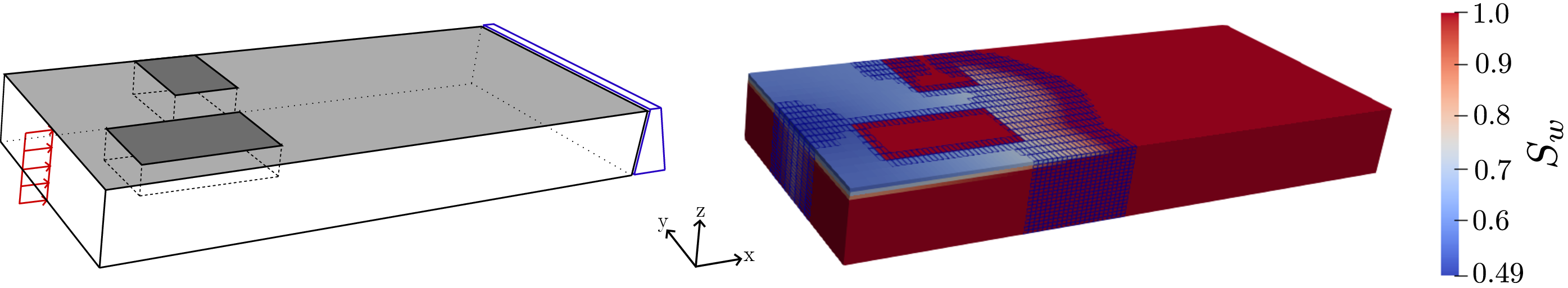}
        \caption{\revision{Hybrid, adaptively coupled injection test. Two impermeable lenses (dark grey) at the top. 120x20x20 grid elements. Domain height $H=\qty{30}{\meter}$}}
        \label{fig_case_c}
    \end{subfigure}
    \caption{\revision{Boundary conditions, geometries and solution fields for three test cases. The first column depicts boundary conditions and impermeable lenses, while the second column illustrates the saturation fields at the end of the simulations. For the hybrid models (\cref{fig_case_b,fig_case_c}), the FD subdomain is denoted via element edges in the solution field, while the smooth surface represents the VE subdomain}}
    \label{fig_all_cases}
\end{figure}

\revision{To evaluate the potential speedup achievable through data-driven methods, we investigate three test cases, that are depicted in \cref{fig_all_cases}. The first test case constitutes an injection scenario which is modelled using only the VE model. The second test incorporates an impermeable lens in the flow field, and is simulated using a static, hybrid approach which couples the FD and VE model. Here, the coupling interface is chosen manually and remains static throughout the simulation. Finally, the third test places two impermeable lenses into the flow field and is simulated with a adaptive, hybrid approach. The adaptive criteria \cite{buntic2025} \revisionTwo{}{for the domain decomposition} are based on error minimization and mass exchange thresholds and are computed per grid column. For the adaptive test case, the coupling interface is determined automatically during runtime and adapted during each time step.}

\revision{In all three test scenarios, methane is injected into a saturated aquifer with different aquifer heights. We adopt an isothermal, immiscible, compressible two-phase formulation using the Darcy equations for momentum conservation and mass balance equations as described in \cref{ch_math_models}. The wetting phase is represented by water as a proxy for brine, while methane serves as the non-wetting phase. Methane is modelled using the ideal gas law, whereas the properties of the wetting phase are computed according to the IAPWS formulation for water~\cite{IAPWS1997}, with the temperature being fixed at \qty{53}{\celsius}. The computational domain is a structured 3D Cartesian grid. Methane is injected through a source line spanning the vertical height, see red line source in \cref{fig_all_cases}, at a rate of \qty{1.46e-3}{\kilogram\per\second\per\meter} over approximately 23 days. In all test cases, a hydrostatic pressure distribution and $S_n=0$ are enforced as Dirichlet boundary conditions at the right boundary, see blue profile in \cref{fig_all_cases}. All other boundaries are treated as Neumann no-flow conditions. The homogeneous bulk is modelled with permeability $k=\qty{2e-12}{\square\meter}$ and porosity $\Phi=0.2$ while the Brooks–Corey capillary pressure–saturation relation is \revision{parametrized} with $\lambda=2.0$, entry pressure $p_e=\qty{1e5}{\pascal}$, and zero residual saturations. The impermeable lenses are modelled with a reduced permeability of $k_l=\qty{2e-17}{\square\meter}$ and are \qty{15}{\meter} high in each test. For all injection tests the computational domain shares a length of \qty{250}{\meter} and a width of \qty{100}{\meter} while the domain height varies between, \qty{80}{\meter}, \qty{50}{\meter} and \qty{30}{\meter} among the test cases. The zero-flux condition at the top boundary mimics a caprock, preventing upward escape of injected gas.}

Due to buoyancy, methane is expected to rise near the well, accumulate beneath the caprock, and propagate laterally in the $x$- and $y$-direction. Upon reaching an impermeable lens, the gas front is anticipated to split, flow around the lens, and merge downstream in its wake. The simulation outcomes with no data-driven integration are illustrated in the right column of \cref{fig_all_cases}, with the initial \revisionTwo{domain decomposition}{conditions} shown on the left. Additionally, the grid refinement varies across the test cases and is recorded in \cref{fig_all_cases}. A finite-volume scheme with an upwinded two-point flux approximation is applied to discretize the governing equations. 

\revision{Using these three test cases as examples, we intend to evaluate the potential speedup achievable through data-driven methods. To this end, we identify three algorithmic blocks for substitution, as shown in \cref{table_comp_effort_all_cases}. For the pure VE model, the computation of the coarse-level mobilities claims the majority of the computational effort (\cref{table_comp_effort_case_a}). In contrary, for the hybrid models (\cref{table_comp_effort_case_b,table_comp_effort_case_c}), the computational bottleneck shifts to the coupling fluxes. Here, the high cost of the coupling fluxes can be backtraced to the repetitive evaluation of secondary variables on the lowest level of the scheme, specifically the density and viscosity. Simultaneously, for the adaptive case (\cref{table_comp_effort_case_c}), the computational effort for the gas plume distance becomes non-negligible, since additional evaluations are required for the computation of the adaptivity criteria. While the gas plume distance calculation is not a major computational bottleneck in either model, it provides a suitable example for demonstrating how data-driven approaches can improve the convergence behaviour of non-linear solvers. Altogether, we select the data-driven replacements of the gas plume distance, the coarse-level mobilities, and the secondary variables as the primary targets of interest in this work.}

\begin{table}[h!]
    \centering
    \begin{subtable}{\textwidth}
        \centering
        \begin{tabular}{||c | c | c | c||} 
            \hline
            Algorithmic block & Computation time [s] & Fraction of total time [\%]\\
            \hline\hline
            Total & 367.5 & 100.00 \\ 
            \hline
            Gas plume distance & 1.6 & 0.43 \\ 
            \hline
            Coarse-level mobilities & 192.7 & 52.4 \\ 
            \hline
        \end{tabular}
        \caption{\revision{Pure VE injection test}}
        \label{table_comp_effort_case_a}
    \end{subtable}

    \vspace{1.5em}

    \begin{subtable}{\textwidth}
        \centering
        \begin{tabular}{||c | c | c | c||} 
            \hline
            Algorithmic block & Computation time [s] & Fraction of total time [\%]\\
            \hline\hline
            Total & 734.8 & 100.00 \\ 
            \hline
            Gas plume distance & 11.5 & 1.6 \\ 
            \hline
            Coarse-level mobilities & 51.3 & 7.0 \\ 
            \hline
            Coupling fluxes & 406.2 & 55.3 \\ 
            \hline
        \end{tabular}
        \caption{\revision{Hybrid, statically-coupled injection test}}
        \label{table_comp_effort_case_b}
    \end{subtable}

    \vspace{1.5em}

    \begin{subtable}{\textwidth}
        \centering
        \begin{tabular}{||c | c | c | c||} 
            \hline
            Algorithmic block & Computation time [s] & Fraction of total time [\%]\\
            \hline\hline
            Total & 2255.9 & 100.00 \\ 
            \hline
            Gas plume distance & 146.1 & 6.5 \\ 
            \hline
            Coarse-level mobilities & 128.4 & 5.7 \\ 
            \hline
            Coupling fluxes & 1374.4 & 60.1 \\ 
            \hline
        \end{tabular}
        \caption{\revision{Hybrid, adaptively-coupled injection test}}
        \label{table_comp_effort_case_c}
    \end{subtable}
    \caption{\revision{Comparison of the computational effort for different algorithmic blocks of the models utilized in \cref{fig_all_cases}}}
    \label{table_comp_effort_all_cases}
\end{table}

As a simulation framework for flow in porous media, we employ {DuMu\textsuperscript{x}}~\cite{dumux38,koch21}, which is implemented in C++ as a submodule of the DUNE numerics library~\cite{dune21}. For the machine learning backend, we rely on scikit-learn~\cite{scikit2011} in Python and onnx runtime~\cite{onnx2021} in C++, while bicubic spline interpolation is supported by the GSL library~\cite{gsl} in C++. All simulations were executed sequentially on an AMD Ryzen 7 4700 CPU with eight cores at a maximum clock speed of 2000 MHz and 14.8 GiB RAM.

\subsection{Enhancing the gas plume distance} \label{ch_predict_zp}
In this section, we demonstrate how data-driven approaches can be employed to enhance the computation of the gas plume distance. As discussed in \cref{ch_VE_model}, the gas plume distance is determined by solving a non-linear equation that enforces mass balance within a grid column. Although this computation is not a major contributor to runtime, see \cref{table_comp_effort_all_cases}, it serves as a representative case study illustrating how data-driven methods can meaningfully accelerate non-linear solvers and may offer broader implications for similar problems.

In our implementation, the gas plume distance $z_p$ is obtained by utilizing a Newton–Raphson method to solve \cref{eq_gasPlumeDist_integral_extended}. A key opportunity for improvement lies in predicting a more accurate initial guess of the solver, thereby reducing the number of Newton iterations that are required for convergence. Replacing the entire computation with a data-driven surrogate is avoided, since even small deviations from the correct value compromise mass conservation and can cause convergence issues for the global system matrix. 

\begin{figure}[h!]
    \begin{center}
        \includegraphics[width=1.0\textwidth]{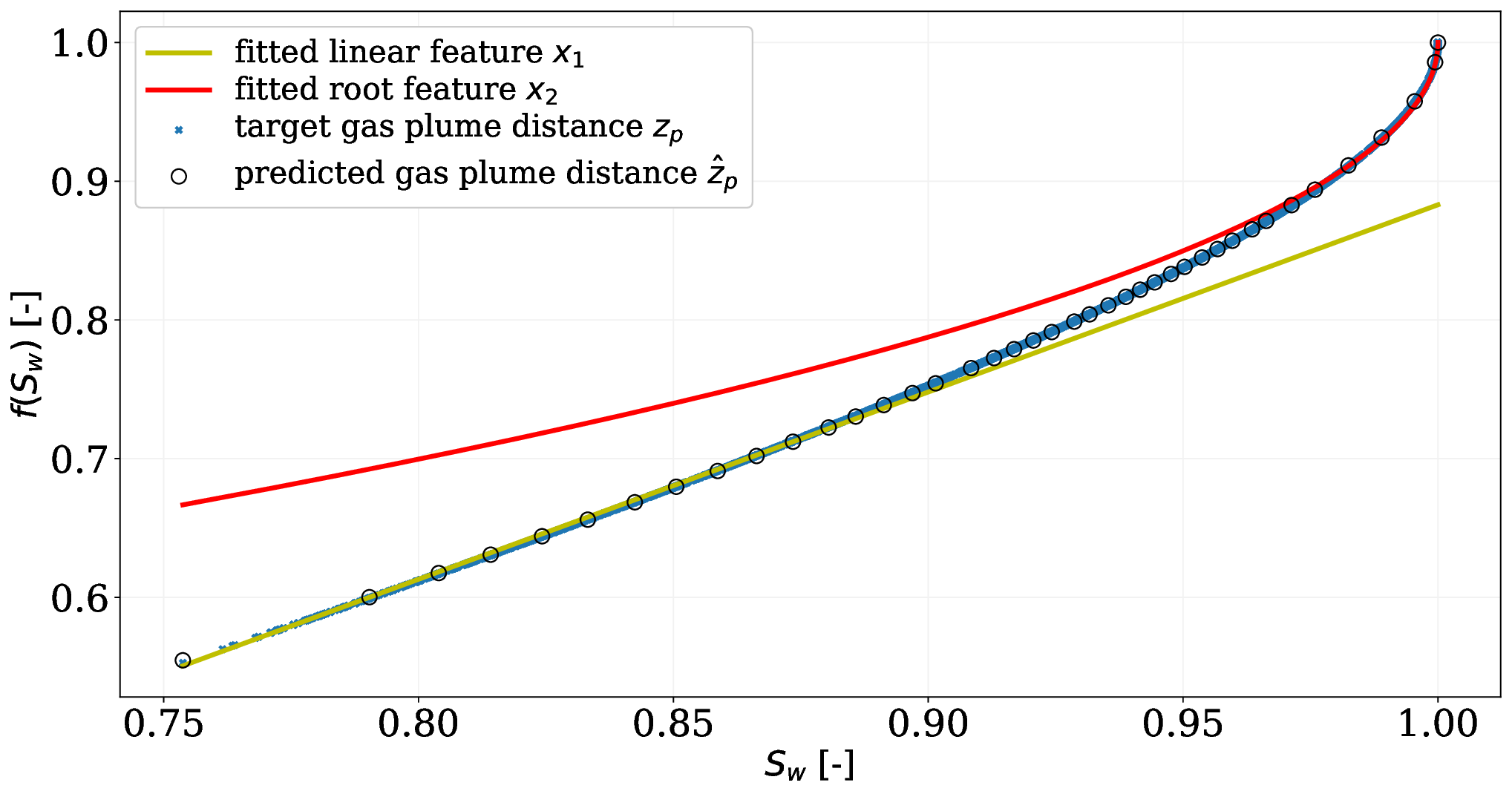}
    \end{center}
    \caption{\revision{Target} relative gas plume distance and proposed saturation-based features. The relative plume distance is defined as the gas plume distance normalized by the domain height \revision{$H=z_T-z_B$}\revisionTwo{}{. The x-axis displays the wetting-phase saturation $S_w$, while the y-axis displays various functions $f(S_w)$ of the wetting-phase saturation, namely the fitted linear and root features $x_1(S_w)$ and $x_2(S_w)$ as well as the predicted and observed gas plume distance}}
    \label{fig_zp_features}
\end{figure}

For this purpose, we select linear regression, which is computationally inexpensive \revision{and} straightforward to implement. The most critical step in achieving satisfactory predictions is the selection of representative features, which is typically an iterative process. If the governing equation is available, potential features can be inferred from the quantities it contains. Parameters that remain constant throughout a simulation can be \revision{excluded from the pool of potential features}. Following this reasoning, and inspired by \cref{eq_gasPlumeDist_integral_extended}, we kept the porous-medium properties $p_e$, $z_B$, $z_T$, $s_{w,r}$, $s_{n,r}$, and $g_z$ fixed, leaving $\lambda$, $(\varrho_w - \varrho_n)$ and $S_w$ as feature candidates. \revision{Strictly speaking, $\lambda$ remains constant during a simulation but showed a considerable impact on the gas plume distance when varied. Through feature-selection methods, we identified $S_w$ and $\lambda$ as the main features that impact the solution of the gas plume distance. The blue data points in \cref{fig_zp_features} represent the target which can be separated into two distinct regimes: for lower wetting-phase saturations, the gas plume distance exhibits an almost linear trend, while for saturations approaching one, the curve resembles a shifted negative square-root function. The linear contribution, represented by the yellowish line in \cref{fig_zp_features}, only depends on the saturation, while the curved tip of the graph is controlled by a non-linear function of the saturation and $\lambda$. Ultimately, we identified the two features} 
\begin{align}
    x_1 &= S_w \label{eq_zpx1}, \\
    x_2 &= -\sqrt{\frac{1}{\lambda}(1.0-S_w)} + 1.0, \label{eq_zpx2}
\end{align}
\revision{which correspond to the linear and non-linear contributions in \cref{fig_zp_features}. The training data was generated using weighted Latin hypercube sampling with the data density being skewed towards $S_w=1$. For \cref{fig_zp_features}, $\lambda=2$ was kept constant, while $S_w$ was represented by 2000 data points. Also note, that we predict a relative gas plume distance, which is the gas plume distance normalized by the domain height. Thereby, our prediction is invariant to the domain height and can be applied to arbitrary domains.}

The accuracy of the surrogate model can be assessed using the coefficient of determination $R^2$, which measures how well the observed data are replicated by the model. Using $x_1$ and $x_2$ as features yielded $R^2$ values close to one, indicating that the selected features accurately capture the behaviour of the gas plume distance, though this does not exclude the existence of superior alternatives. Applying linear regression to these features resulted in the fitted relation
\begin{equation}
    \hat{z}_p = -0.4573 + 0.6301 x_1 + 0.8271 x_2,
\end{equation}
where $\hat{z}_p$ denotes the predicted initial value of the relative gas plume distance. \revision{For comparison, in the original scheme, half the domain height was assigned as the initial guess.}

\revision{As a result, the number of required iterations until convergence could be reduced from 8.5 to 4.8 the pure VE case (\cref{fig_case_a}), from 6.8 to 4.1 for the statically coupled case (\cref{fig_case_b}) and from 10.5 to 4.7 for the adaptively coupled case (\cref{fig_case_c}). In more complex problems, where local non-linear solves account for a larger share of runtime, such optimizations can yield substantial benefits. In comparison, if the gas plume distance from the last time step is used as the initial guess, the iterations are also reduced but not as significantly as in the case of the data-driven approach while coming with the overhead of tracking values across time steps. As evident in \cref{table_solver_statistics_all_cases}, the behaviour of the global solver does not change significantly when replacing the computation of $z_p$ by a surrogate. While no error is introduced in the solution fields for the first two cases (\cref{table_rel_errors_case_a,table_rel_errors_case_b}) since we only approximate the initial prediction of $z_p$, the adaptive case exhibits minor deviations from the reference solution (\cref{table_rel_errors_case_c}). This effect arises from minor discrepancies in the adaptive partitioning of the domain. Because the numerical solution for $z_p$ is computed only up to a residual tolerance of \num{1e-8}, the adaptivity criteria that depend on $z_p$ may occasionally trigger different partitioning decisions. Consequently, small errors appear in those columns where the reference solution and the surrogate-based solution employ different local models.}

To quantify accuracy, we compared the surrogate-enhanced solutions to the original ones using the relative $\mathrm{L}_2$ error:

\begin{equation}
    \mathrm{L}_2(u) = 
    \begin{cases}
        \frac{\sqrt{\sum_i (u_i-u_{m,i})^2}}{\sqrt{\sum_i u_i^2}}                ,& \text{if } \sum_i u_i^2 > 0\\[1.5em] 
        \sqrt{\sum_i u_{m,i}^2},& \text{if } \sum_i u_i^2 = 0,
    \end{cases}
\end{equation}
where $u$ denotes the reference solution, $u_m$ the surrogate-enhanced solution, and the sums extend over all grid elements $i$. In addition, we evaluated the relative non-wetting-phase mass balance error,
\begin{equation}
    m_{n,e} = \frac{m_n - m_{n,s}}{m_n},
\end{equation}
with $m_n$ the expected non-wetting-phase mass and $m_{n,s}$ the simulated mass. \revision{Accordingly, the error for all solution fields of all three test cases is documented in \cref{table_rel_error_all_cases}.}

\subsection{Surrogate for the coarse-level mobilities} \label{ch_predict_coarse_mobilities}
In contrast to the gas plume distance, the computation of coarse-level mobilities is considerably more resource-intensive, particularly in the pure VE model, see \cref{table_comp_effort_case_a}. Optimizing this algorithmic block therefore promises greater benefits. As indicated in \cref{eq_integratedQuantities}, the coarse-level mobilities are obtained by vertically integrating weighted fine-level mobilities. In discrete form, this integration is realized as a summation over all fine-level elements within a column, which implies that the computational cost scales directly with the number of vertical grid cells. To reduce this overhead, we aim to replace the computation of coarse-level mobilities with data-driven surrogates.
\begin{figure}[h!]
    \begin{center}
        \includegraphics[width=1.0\textwidth]{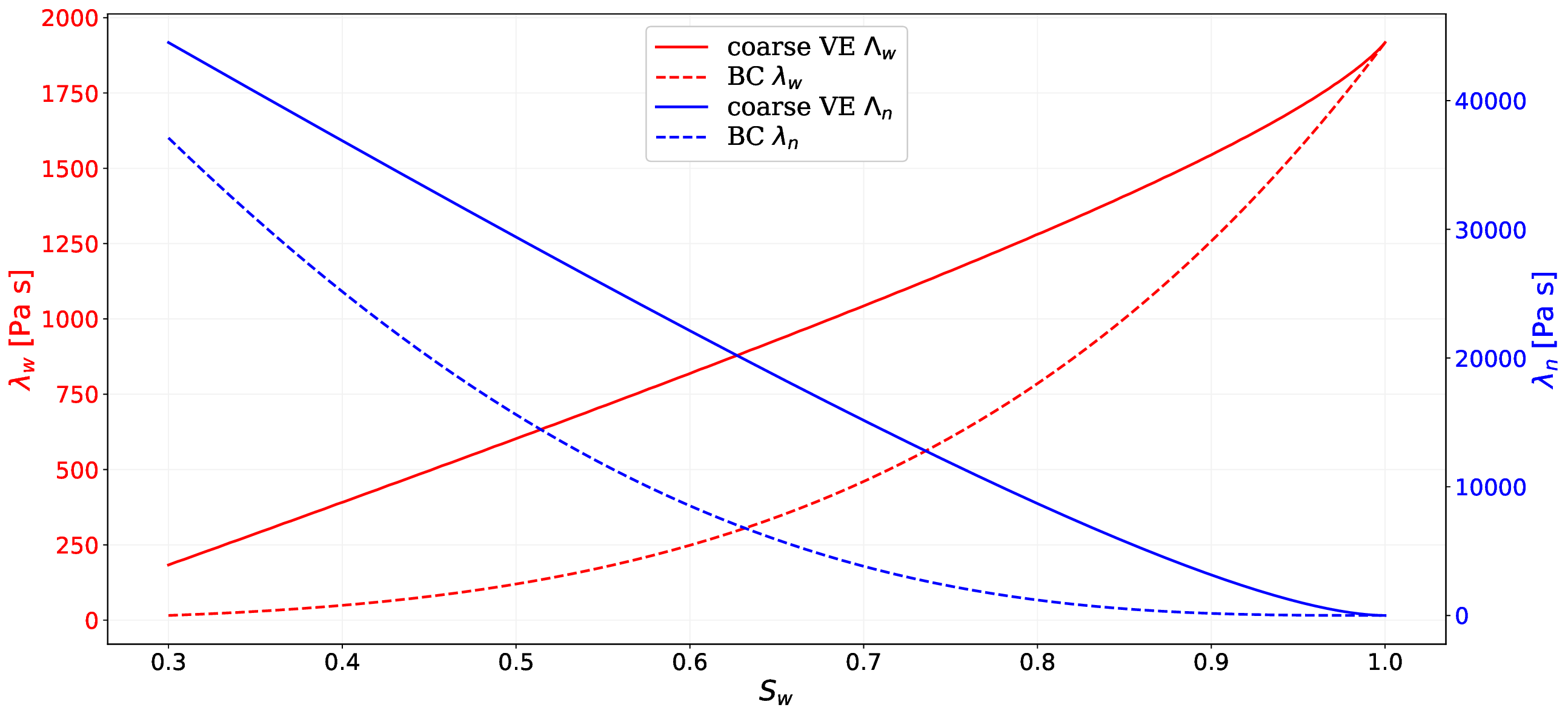}
    \end{center}
    \caption{Comparison of target coarse-level mobilities with Brooks–Corey mobilities computed using fixed $\lambda=2$}
    \label{fig_mobilities}
\end{figure}
In \cref{fig_mobilities}, the solid curves represent the true coarse-level VE mobilities obtained by vertical integration while the dashed curves show mobilities that are computed by directly applying the Brooks–Corey relations to the coarse-level saturation \revision{for a} fixed uniformity coefficient $\lambda = 2$. \revision{It is notable, that the coarse-scale mobilities exhibit a linear regime for smaller saturations and transfer to a slightly non-linear relation for saturations close to one, which resembles the behaviour of the gas plume distance in \cref{fig_zp_features}.}

\revision{Initially}, we investigated substituting the vertical integration step with linear regression models. Inspired by \cref{eq_relPermSat_w,eq_phaseMobility}, we considered the wetting-phase saturation as the dominant feature. \revision{The feature space was expanded via polynomial feature expansion and reduced to the most significant polynomial contributions via dimensionality-reduction techniques. However, the linear regression models were not accurate enough and also yielded non-monotone predictions, which compromised numerical stability and mass conservation.} For mobility functions, monotonicity and continuity are two essential properties. Monotonicity ensures numerical stability of the resulting fluxes, while continuity guarantees mass conservation across the flow field. To satisfy both requirements, spline interpolation emerged as a promising alternative. Cubic basis functions allow smooth, continuous transitions across patches and can be tuned to preserve monotonic behaviour. Provided that the training data adequately cover the feature space, spline interpolation can deliver near-exact reproductions \revision{of smooth functions}. \revision{Again, motivated by the expanded formulation of the coarse-level mobility in \cref{eq_integratedQuantities}, we identified}
\begin{align}
    x_3 &= S_w \label{eq_mobx3} \\
    x_4 &= \varrho_w \label{eq_mobx4}
\end{align}
\revision{as the most significant features, with $\varrho_w$ acting as a proxy for the pressure. While the pressure can range several logarithmic scales, the density remains on the same scale with varying pressure. The training data were generated using a weighted Latin hypercube sampling for the saturation and a regular Latin hypercube sampling for the density. As a result, the saturation space was populated with 2000 points skewed towards $S_w=1.0$ for the range \qtyrange{0.3}{1.0}{} while the density space was covered by 60 points for the range \qtyrange{987.0}{1000.0}{\kilo\gram\per\meter\cubed}. We did not cover the range \qtyrange{0.0}{0.3}{} for $S_w$ since these values were not relevant for our simulations and allowed deploying more training data for the relevant ranges. Since our test cases are conducted under isothermal conditions, the wetting-phase pressure can be extracted from the density values by applying the inverse density function, thus allowing the computation of the target coarse-level mobilities given the saturation and pressure.} Introducing additional features was deliberately avoided, as the computational cost of spline interpolation grows exponentially with the dimensionality of the input space. Explicit coefficients for $x_3$ and $x_4$ cannot be given globally, since they are determined individually within each patch of the spline mesh.

It is important to note that the surrogate predictions for mobilities are only valid under the assumption of homogeneous permeability distributions within the VE grid columns, consistent with the setup used to generate the training data. As evident from \cref{eq_integratedQuantities}, permeability distributions directly influence coarse-level mobilities. However, the classical VE model itself is primarily suited to homogeneous or near-homogeneous regions, and should not be deployed in heterogeneous zones. While extensions of VE models that account for heterogeneous layering exist~\cite{guo2016,moyner2019}, such variants are not employed in this study.

Using $x_3$ and $x_4$ as features, we constructed spline interpolators for the wetting- and non-wetting-phase coarse-level mobilities, $\hat{\Lambda}_w(x_3,x_4)$ and $\hat{\Lambda}_n(x_3,x_4)$. Interestingly, even though both features are defined by wetting-phase quantities, they also proved effective in predicting non-wetting-phase mobilities. 

\revision{Utilizing the mobility surrogate, for the pure VE test, the computational cost of mobility evaluations was reduced by more than 99\%. Similarly, the mobility computation exhibited a reduction of 99\% for the statically-coupled test and 98\% for the adaptively coupled test. As in the gas plume distance enhancement, the induced error in mass balance was negligible.}

\revision{Overall, introducing a surrogate for the coarse-level mobility introduces minor, acceptable errors in the solution field for all three test scenarios (\cref{table_rel_error_all_cases}). Although the errors for non-wetting-phase quantities may be larger, this is an artifact of the relative error computation. Initially, the domain is fully saturated with brine, such that non-wetting-phase quantities start at zero. Relative error measures amplify small deviations near zero, making the non-wetting-phase errors appear large, even when their absolute magnitude is small.}

\revision{Additionally, as presented in \cref{table_solver_statistics_case_a,table_solver_statistics_case_b}, the number of required time steps decreases substantially when coarse-level mobilities are substituted with surrogates. A likely explanation is a smoothing effect introduced by the surrogate. As illustrated in \cref{fig_flowcharts}, the standard computation of coarse-level mobilities involves several stages: computing fine-level saturations and mobilities, averaging them locally to reduce discretization artifacts, and then integrating vertically according to \cref{eq_integratedQuantities}. This sequence can introduce small discontinuities or inconsistencies in the computed mobilities, which reduce solver efficiency. In contrast, a surrogate such as a spline or linear regressor acts as a smoothing filter, suppressing minor fluctuations and yielding more numerically stable mobility functions. This improved smoothness enhances solver performance, allowing larger time steps without increasing the average number of Newton or linear iterations per step.}

In summary, substituting coarse-level mobilities with spline-based surrogates provides a substantial speedup while introducing only minor errors.

\subsection{Enhancing the coupling scheme via surrogate secondary variables} \label{ch_predict_secondaries}
While the pure VE model is significantly accelerated through surrogates for the gas plume distance and coarse-level mobilities, the hybrid scheme still suffers from high computational costs dominated by the evaluation of coupling fluxes. Since the hybrid framework operates with two models of different dimensionality in the same computational domain, coupling fluxes are indispensable for exchanging information across models.

\begin{figure}[h!]
    \begin{center}
        \includegraphics[width=1.0\textwidth]{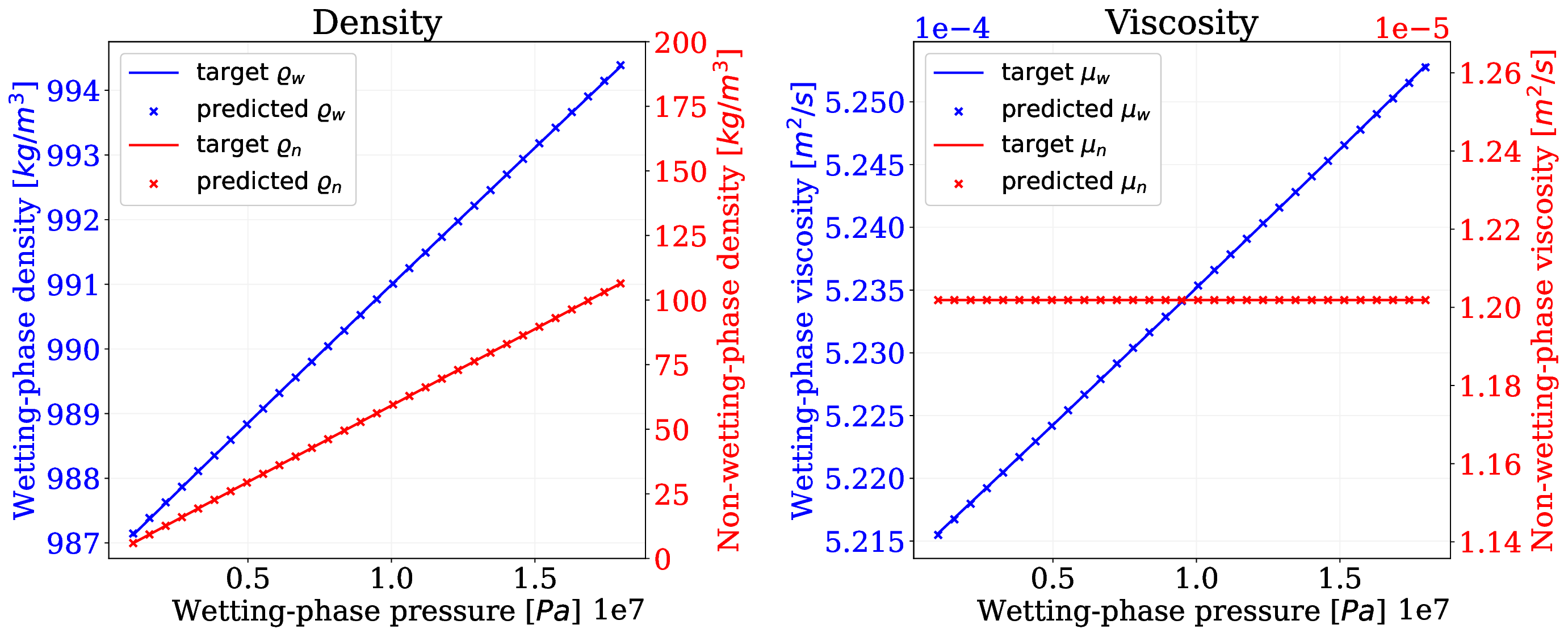}
    \end{center}
    \caption{Comparison of the target densities (left) and target viscosities (right) with the predicted values}
    \label{fig_secondaries}
\end{figure}

A more careful examination of the flux computation bottleneck reveals that the principal cost arises not from the flux evaluation itself but from the computation of secondary variables, specifically phase densities and viscosities. Since the non-wetting phase\revision{, in our case methane,} is modelled as an ideal gas, its secondary variables are inexpensive to compute. By contrast, the wetting-phase quantities rely on the accurate but computationally intensive IAPWS relations~\cite{IAPWS1997}. Because these evaluations are invoked repeatedly for each element in every Newton iteration, the cumulative effort becomes substantial. The computation of coupling fluxes, in particular, demands a large number of such evaluations. To address this bottleneck, we shifted focus to constructing data-driven surrogates for density and viscosity. The main inputs for these quantities are pressure and temperature. However, as our model is isothermal, temperature can be neglected, leaving the wetting-phase pressure as the sole feature,
\begin{equation}
    x_5 = p_w. \label{eq_mobx5}
\end{equation}
\revision{The training data was generated using a classical Latin hypercube sampling method where the range \qtyrange{1e6}{1.8e7}{\pascal}, which is representative for our test cases, was covered by 10000 data points.} Including higher-order contributions of the pressure, such as squared or cubic terms, did not lead to any meaningful improvement in predictive accuracy. After training the linear regression models, we obtained the following predictive relations:
\begin{alignat}{3}
    \hat{\varrho}_w &= 986.7185 &&+ 4.2648\text{e}^{-7} &&p_w,\\
    \hat{\varrho}_n &= &&\quad\: 5.9188\text{e}^{-6} &&p_w,\\
    \hat{\mu}_w &= 5.2133\text{e}^{-4} &&+ 2.1941\text{e}^{-13} &&p_w,\\
    \hat{\mu}_n &= 1.2019\text{e}^{-5}, && &&
\end{alignat}
where $\hat{\varrho}_w$ and $\hat{\varrho}_n$ denote the predicted wetting-phase and non-wetting-phase densities, respectively, while $\hat{\mu}_w$ and $\hat{\mu}_n$ denote the corresponding viscosities. The gaseous phase viscosity remains constant in our setup, as the viscosity of an ideal gas depends only on temperature and not on pressure~\cite{reid1987}. As illustrated in \cref{fig_secondaries}, the surrogates provide an almost perfect match to the target values, achieving $R^2$ values effectively equal to one. This strong agreement is expected, as both densities and viscosities display either linear or constant trends within the pressure range relevant to our study, making linear regression particularly suitable.

By replacing the evaluation of densities and viscosities with these surrogates only within the coupling scheme, we achieve a computational speedup of 22\% for the static test scenario and 17.6\revisionTwo{}{\%} for the adaptive one. Although this percentage appears modest compared to the improvements from previous surrogates, it translates into a substantial reduction in the total simulation runtime, given that the coupling scheme dominates the computational workload.

To further enhance efficiency, we extended the use of the surrogate model to the entire simulation, substituting all evaluations of densities and viscosities. The error analysis in \cref{table_rel_error_all_cases} confirms that the relative errors introduced by this substitution remain consistently below one percent.

Finally, according to the runtime statistics reported in \cref{table_solver_statistics_all_cases}, the use of surrogates for secondary variables does not significantly alter the non-linear or linear solver behaviour, indicating that the substitution preserves numerical stability while substantially reducing computational cost.

\subsection{Combined deployment of data-driven surrogates} \label{ch_combined_surrogates}
Finally, we analyse the scenario in which all presented surrogate models are applied simultaneously, see column 4) in \cref{table_rel_error_all_cases,table_solver_statistics_all_cases} for reference. For the pure VE model, the total computation time is reduced by approximately 75.4\%, while for the static coupling approach the reduction amounts to 44.4\% and for the adaptive case to 18\%. The rather low increase for the adaptive case can be attributed to the dominating share of the FD subdomain to the total domain. At the same time, the structure of the global system matrix appears unaffected since the average number of linear solver iterations per Newton step, as well as the average number of Newton iterations per time step, remains nearly unchanged. Furthermore, the overall scheme preserves mass conservation while producing solution fields that deviate only negligibly from the reference results. Importantly, surrogate models such as those for the gas plume distance or secondary variables enhance, but do not replace, underlying physical processes. This allows for greater flexibility, since these surrogates can be deployed robustly under different initial and boundary conditions without retraining.

A direct comparison with a pure FD simulation highlights the benefits of this approach. For the adaptively, coupled setup shown in \cref{fig_case_c}, the unoptimized hybrid model required roughly \qty{2256}{\second}, whereas the corresponding pure FD simulation finished in about \qty{1976}{\second}. After incorporating the data-driven surrogates, however, the hybrid model runtime dropped to approximately \qty{1894}{\second}. This demonstrates that the optimized hybrid model not only outperforms its own unoptimized counterpart but also becomes faster than the full FD simulation while still maintaining physical fidelity. The achievable speedup naturally scales with the complexity of the coupling interface and the ratio of VE to FD elements with smaller ratios yielding smaller gains, while more complex configurations benefit more strongly.

Solver statistics provide further insight. For the pure FD simulation, a total of 261 Newton iterations were required, with an average of 77.69 linear solver iterations per Newton step. In contrast, the enhanced adaptive model converged in 268 Newton iterations, but with only 56.74 linear solver iterations per Newton step on average. This indicates that, despite requiring more Newton iterations, the hybrid system matrix is structurally more efficient to solve. At first glance, this may appear counter-intuitive, since coupling terms introduce off-diagonal entries that generally reduce solver efficiency. However, the hybrid formulation also reduces the total number of degrees of freedom, which in practice dominates and improves overall efficiency. Consequently, the hybrid model with surrogates achieves shorter runtimes even though it requires more Newton steps.

\begin{figure}[h!]
    \centering
    \begin{subfigure}{\textwidth}
        \centering
        \includegraphics[width=\textwidth]{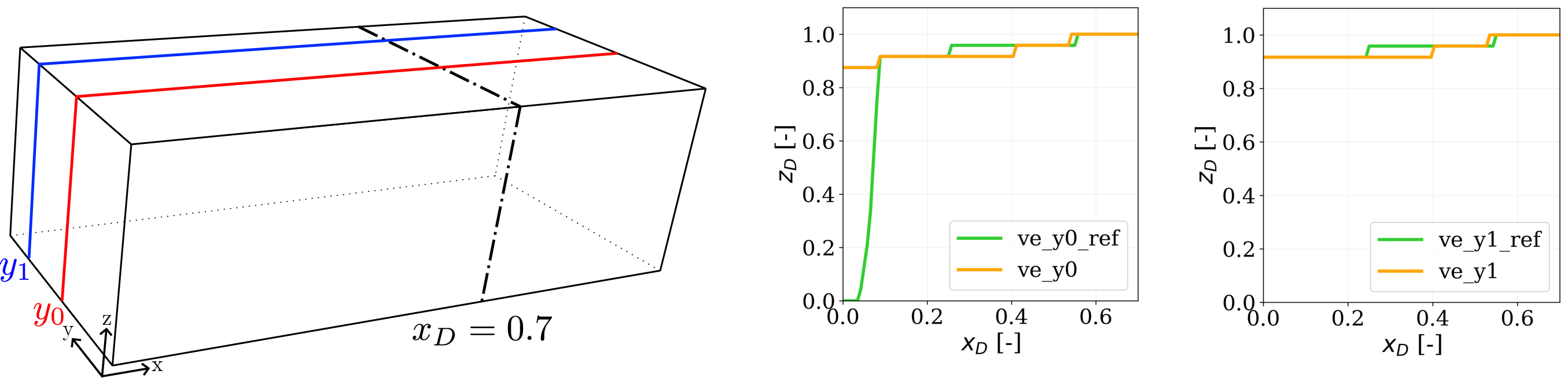}
        \caption{\revisionTwo{}{Pure VE injection test}}
        \label{fig7_case_a}
    \end{subfigure}

    \vspace{1em}

    \begin{subfigure}{\textwidth}
        \centering
        \includegraphics[width=\textwidth]{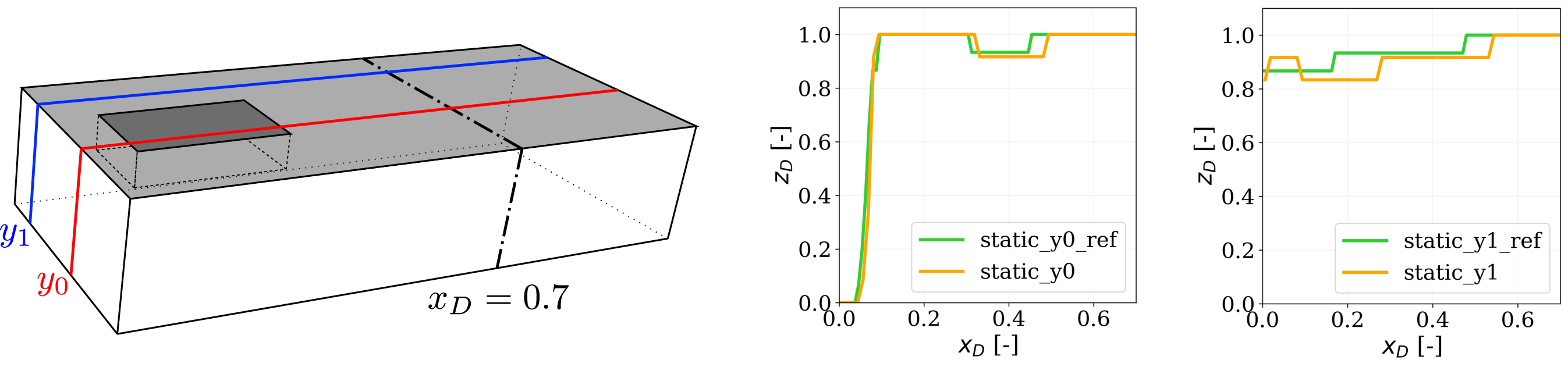}
        \caption{\revisionTwo{}{Hybrid, statically-coupled injection test}}
        \label{fig7_case_b}
    \end{subfigure}

    \vspace{1em}

    \begin{subfigure}{\textwidth}
        \centering
        \includegraphics[width=\textwidth]{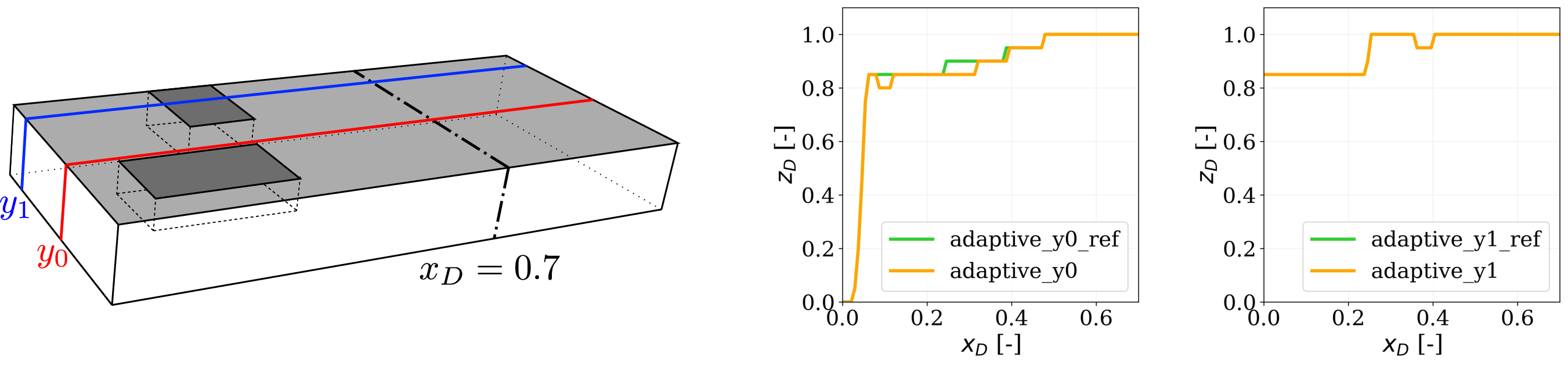}
        \caption{\revisionTwo{}{Hybrid, adaptively coupled injection test}}
        \label{fig7_case_c}
    \end{subfigure}
    \caption{\revisionTwo{}{Comparison of the gas plume shape at the end of the simulation between the reference FD solution (green) and the fully-surrogated VE models (orange). The plume shape is inspected along x-z-slices at two different, constant depths $y_0=50\text{m}$ and $y_1=83\text{m}$. For simplicity, the graphs stop at the dimensionless length $x_D=0.7$, since the plume does not exceed that length in any of the simulations. The graphs depict a dimensionless length $x_D$ on the x-axis, and a dimensionless height $z_D$ on the y-axis which are obtained by dividing the x-position by the length of the domain and the z-position by the height of the domain. The left graph belongs to $y_0$ while the right graph belongs to $y_1$}}
    \label{fig_plume_comparison}
\end{figure}

\revisionTwo{}{Additionally, in \cref{fig_plume_comparison}, we compare the spatial distribution of the gas plume at the end of the simulation. To this end, we illustrate the gas plume shape along two different slices at constant depths $y_0$ and $y_1$. As evident in \cref{fig7_case_a}, the pure VE model fails to capture the vertical motion of the gas around the injection well, and instead assumes a state of phase segregation. Yet, the predicted plume shape at $y_1$ matches well with the reference solution. Comparing the plume shapes of the adaptively and statically coupled models in \cref{fig7_case_b} and \cref{fig7_case_c}, we observe that the adaptive model showcases overall better predictive capabilities since the it adaptively captures dynamic processes via the FD model while employing the VE model only where accuracy allows it.}

\begin{figure}[h!]
    \begin{center}
        \includegraphics[width=0.8\textwidth]{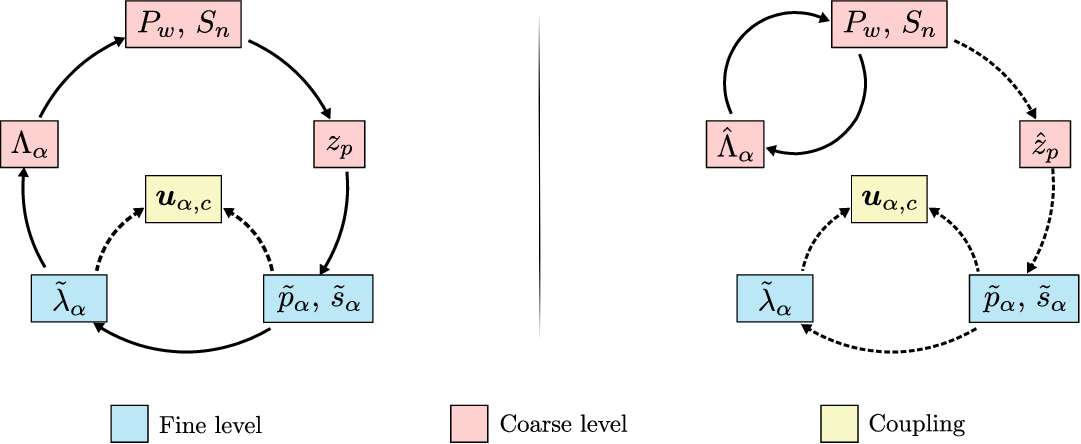}
    \end{center}
    \caption{Flowcharts depicting the solving strategy of the pure VE model (solid lines) and hybrid model (solid+dashed lines). The left flowchart describes the strategy when no data-driven models are utilized, while the right one incorporates enhancement by data-driven approaches. Red boxes indicate the computation on the coarse level of the VE model while blue ones indicate computations on the fine level and yellow boxes represent operations necessary for computing the coupling fluxes in the hybrid approach}
    \label{fig_flowcharts}
\end{figure}

The use of surrogates also streamlines the solution cycle of the VE model. As illustrated in \cref{fig_flowcharts}, the left flowchart shows the baseline pure VE scheme, where each iteration requires the computation of the gas plume distance $z_p$, reconstruction of pressures, saturations, and fine-level mobilities, followed by the computation of the coarse-level mobilities. For the hybrid model, an additional step is required, namely the evaluation of coupling fluxes using reconstructed quantities, as denoted by the dashed lines. The right flowchart, in contrast, illustrates the surrogate-augmented scheme. Here, the pure VE model reduces to two main steps. First, the prediction of the coarse-level mobilities $\hat{\Lambda}_\alpha$ from the primary variables, and second, the computation of the primary variables using the predicted mobilities. For hybrid simulations, fine-level VE quantities are still required for visualizations and coupling flux computations, but surrogates for the gas plume distance $\hat{z}_p$ and the secondary variables, specifically densities and viscosities, can be exploited within each iteration.

\begin{table}[h!]
    \centering
    \begin{subtable}{\textwidth}
        \centering
        \begin{tabular}{||c | c | c | c | c||}
            \hline
             & 1) & 2) & 3) & 4)\\ 
            \hline\hline
            $\mathrm{L}_2(S_w)$ & 0.00e+00 & 3.11e-03 & 2.13e-07 & 3.11e-03\\ 
            \hline
            $\mathrm{L}_2(S_n)$ & 0.00e+00 & 2.23e-01 & 5.01e-06 & 2.23e-01\\ 
            \hline
            $\mathrm{L}_2(P_w)$ & 0.00e+00 & 7.11e-05 & 3.10e-07 & 7.11e-05\\ 
            \hline
            $\mathrm{L}_2(P_n)$ & 0.00e+00 & 8.26e-04 & 9.58e-07 & 8.26e-04\\ 
            \hline
            $\mathrm{L}_2(\Lambda_w)$ & 0.00e+00 & 1.17e-02 & 1.31e-05 & 1.17e-02\\ 
            \hline
            $\mathrm{L}_2(\Lambda_n)$ & 0.00e+00 & 3.64e-01 & 4.28e-06 & 3.64e-01\\ 
            \hline
            $\mathrm{L}_2(\varrho_w)$ & 0.00e+00 & 5.19e-07 & 1.07e-05 & 1.04e-05\\ 
            \hline
            $\mathrm{L}_2(\varrho_n)$ & 0.00e+00 & 7.10e-05 & 1.33e-07 & 7.10e-05\\ 
            \hline
            $m_{n,e}$ & 2.27e-13 & 7.40e-13 & -2.36e-14 & -3.98e-13\\ 
            \hline
        \end{tabular}
        \caption{\revision{Pure VE injection test}}
        \label{table_rel_errors_case_a}
    \end{subtable}

    \vspace{1.5em}

    \begin{subtable}{\textwidth}
        \centering
        \begin{tabular}{||c | c | c | c | c | c||}
            \hline
             & 1) & 2) & 3*) & 3) & 4)\\ 
            \hline\hline
            $\mathrm{L}_2(S_w)$ & 0.00e+00 & 1.19e-03 & 1.36e-06 & 8.09e-05 & 1.20e-03\\ 
            \hline
            $\mathrm{L}_2(S_n)$ & 0.00e+00 & 8.83e-02 & 9.97e-05 & 6.02e-03 & 8.91e-02\\ 
            \hline
            $\mathrm{L}_2(P_w)$ & 0.00e+00 & 2.40e-05 & 9.27e-08 & 1.56e-06 & 2.49e-05\\ 
            \hline
            $\mathrm{L}_2(P_n)$ & 0.00e+00 & 2.37e-04 & 3.46e-07 & 2.50e-05 & 2.37e-04\\ 
            \hline
            $\mathrm{L}_2(\Lambda_w)$ & 0.00e+00 & 4.84e-03 & 4.34e-06 & 2.81e-04 & 4.93e-03\\ 
            \hline
            $\mathrm{L}_2(\Lambda_n)$ & 0.00e+00 & 2.28e-01 & 1.05e-04 & 5.57e-03 & 2.22e-01\\ 
            \hline
            $\mathrm{L}_2(\varrho_w)$ & 0.00e+00 & 1.83e-07 & 0.00e+00 & 1.08e-05 & 1.07e-05\\ 
            \hline
            $\mathrm{L}_2(\varrho_n)$ & 0.00e+00 & 2.40e-05 & 1.10e-07 & 1.64e-06 & 2.49e-05\\ 
            \hline
            $m_{n,e}$ & 2.00e-16 & 5.99e-16 & -2.00e-16 & 0.00e+00 & 3.99e-16\\ 
            \hline
        \end{tabular}
        \caption{\revision{Hybrid, statically-coupled injection test}}
        \label{table_rel_errors_case_b}
    \end{subtable}

    \vspace{1.5em}

    \begin{subtable}{\textwidth}
        \centering
        \begin{tabular}{||c | c | c | c | c | c||}
            \hline
             & 1) & 2) & 3*) & 3) & 4)\\ 
            \hline\hline
            $\mathrm{L}_2(S_w)$ & 3.89e-06 & 3.91e-06 & 1.48e-06 & 6.38e-05 & 5.98e-05\\ 
            \hline
            $\mathrm{L}_2(S_n)$ & 2.09e-04 & 2.10e-04 & 7.92e-05 & 3.44e-03 & 3.22e-03\\ 
            \hline
            $\mathrm{L}_2(P_w)$ & 5.78e-07 & 5.73e-07 & 1.32e-07 & 4.35e-06 & 4.73e-06\\ 
            \hline
            $\mathrm{L}_2(P_n)$ & 7.25e-07 & 7.24e-07 & 1.95e-07 & 7.63e-06 & 7.84e-06\\ 
            \hline
            $\mathrm{L}_2(\Lambda_w)$ & 7.75e-06 & 1.57e-04 & 2.99e-06 & 1.30e-04 & 1.95e-04\\ 
            \hline
            $\mathrm{L}_2(\Lambda_n)$ & 3.26e-04 & 3.54e-04 & 1.28e-04 & 5.60e-03 & 5.36e-03\\ 
            \hline
            $\mathrm{L}_2(\varrho_w)$ & 3.40e-08 & 3.40e-08 & 0.00e+00 & 1.08e-05 & 1.08e-05\\ 
            \hline
            $\mathrm{L}_2(\varrho_n)$ & 7.24e-07 & 7.25e-07 & 1.85e-07 & 4.41e-06 & 4.77e-06\\ 
            \hline
            $m_{n,e}$ & -4.99e-16 & -1.66e-16 & -4.99e-16 & -1.66e-16 & 1.66e-16\\ 
            \hline
        \end{tabular}
        \caption{\revision{Hybrid, adaptively-coupled injection test}}
        \label{table_rel_errors_case_c}
    \end{subtable}
    \caption{\revision{Relative error analysis of solution fields on the VE coarse level for the test cases in \cref{fig_all_cases}. The original, non-data-driven model serves as the reference solution. Model 1): surrogate gas plume distance, model 2): surrogate coarse-level mobilities, model 3*): surrogate secondaries in flux, model 3): surrogate secondaries everywhere, 4): surrogate gas plume distance, coarse-level mobilities and secondaries everywhere}}
    \label{table_rel_error_all_cases}
\end{table}

\begin{table}[h!]
    \centering
    \begin{subtable}{\textwidth}
        \centering
        \begin{tabular}{||c | c | c | c | c | c||}
            \hline
             & 0) & 1) & 2) & 3) & 4)\\ 
            \hline\hline
            Time steps & 64 & 64 & 49 & 64 & 49\\ 
            \hline
            NSI & 437 & 437 & 275 & 437 & 275\\ 
            \hline
            LSI & 17420 & 17423 & 10656 & 17424 & 10607\\ 
            \hline
            ALPN & 39.86 & 39.87 & 38.75 & 39.87 & 38.57\\ 
            \hline
            ANPT & 6.83 & 6.83 & 5.61 & 6.83 & 5.61\\ 
            \hline
            Runtime [s] & 367.5 & 355.3 & 114.2 & 328.2 & 90.4\\ 
            \hline
        \end{tabular}
        \caption{\revision{Pure VE injection test}}
        \label{table_solver_statistics_case_a}
    \end{subtable}

    \vspace{1.5em}

    \begin{subtable}{\textwidth}
        \centering
        \begin{tabular}{||c | c | c | c | c | c | c||}
            \hline
             & 0) & 1) & 2) & 3*) & 3) & 4)\\ 
            \hline\hline
            Time steps & 56 & 56 & 48 & 56 & 56 & 47\\ 
            \hline
            NSI & 352 & 352 & 262 & 352 & 349 & 260\\ 
            \hline
            LSI & 12656 & 12635 & 9132 & 12631 & 12548 & 9029\\ 
            \hline
            ALPN & 35.95 & 35.89 & 34.85 & 35.88 & 35.95 & 34.73\\ 
            \hline
            ANPT & 6.29 & 6.29 & 5.46 & 6.29 & 6.23 & 5.53\\ 
            \hline
            Runtime [s] & 734.8 & 714.3 & 500.1 & 640.4 & 589.5 & 408.5\\ 
            \hline
        \end{tabular}
        \caption{\revision{Hybrid, statically-coupled injection test}}
        \label{table_solver_statistics_case_b}
    \end{subtable}

    \vspace{1.5em}

    \begin{subtable}{\textwidth}
        \centering
        \begin{tabular}{||c | c | c | c | c | c | c||}
            \hline
             & 0) & 1) & 2) & 3*) & 3) & 4)\\ 
            \hline\hline
            Time steps & 48 & 48 & 48 & 48 & 48 & 48\\ 
            \hline
            NSI & 269 & 269 & 269 & 269 & 266 & 268\\ 
            \hline
            LSI & 15326 & 15269 & 15268 & 15283 & 15221 & 15207\\ 
            \hline
            ALPN & 56.97 & 56.76 & 56.76 & 56.81 & 57.22 & 56.74\\ 
            \hline
            ANPT & 5.60 & 5.60 & 5.60 & 5.60 & 5.54 & 5.58\\ 
            \hline
            Runtime [s] & 2255.9 & 2231.5 & 2172.7 & 2014.7 & 1894.7 & 1849.1\\ 
            \hline
        \end{tabular}   
        \caption{\revision{Hybrid, adaptively-coupled injection test}}
        \label{table_solver_statistics_case_c}
    \end{subtable}
    \caption{\revision{Solver statistics for the test cases in \cref{fig_all_cases}. Model 0): no surrogates, model 1): surrogate gas plume distance, model 2): surrogate coarse-level mobilities, model 3*): surrogate secondaries in flux, model 3): surrogate secondaries everywhere, 4): surrogate gas plume distance, coarse-level mobilities and secondaries everywhere. NSI: \textbf{N}ewton \textbf{s}olver \textbf{i}terations, LSI: \textbf{l}inear \textbf{s}olver \textbf{i}terations, ALPN: \textbf{a}verage number of \textbf{l}inear solver iterations \textbf{p}er \textbf{N}ewton solver iteration, ANPT: \textbf{a}verage number of \textbf{N}ewton iterations \textbf{p}er \textbf{t}imestep}}
    \label{table_solver_statistics_all_cases}
\end{table}

\subsection{Limitations of the presented approaches} \label{ch_limitation}
When employing data-driven surrogates, it is essential to recognize the limitations of the underlying models in order to prevent extrapolation into non-physical regimes \revisionTwo{}{which manifests itself in the non-convergence of local and global solvers}. In the following, we outline the conditions under which each surrogate can be reused and the circumstances that require retraining or adjustments.

For the linear regression surrogate of the gas plume distance, the predicted quantity depends on the chosen fluids. Thus, when simulations involve alternative phase components, the surrogate must be retrained. This retraining requires a new data set that can be produced using the established processes. \revision{Equally, the surrogates for the the mobilities and secondary variables depend on the chosen fluids.}

As previously discussed, the surrogates for coarse-level mobilities inherit the assumption that the VE model is applied exclusively to homogeneous columns. This restriction also applies to the surrogate-enhanced version.

\revision{Modifying the parameters of the Brooks-Corey relations, necessitates an update of the gas plume distance and mobility surrogates. However, the existing linear regression models can be easily updated with the additional data by means of e.g. recursive least squares. On \revisionTwo{}{the} contrary, the spline model needs to be retrained on the accumulated data, which is not computationally expensive as long as the feature space is kept restricted to two features.}

By contrast, the initial and boundary conditions of the simulations can be altered freely without requiring retraining of the surrogates. Similarly, impermeable lenses in the computational domain may be repositioned without adjustment, provided that heterogeneous columns are consistently covered by the FD scheme in the hybrid approach. \revision{As demonstrated by the varied domain height in the test scenarios, all presented surrogate models are invariant to the domain height.}

\section{Conclusion} \label{ch_conclusion}
In this work, we analysed the computational costs and bottlenecks of a pure VE scheme as well as a hybrid approach combining FD and VE models. For the hybrid scheme, we identified the computation of coupling fluxes as the most expensive code block, followed by the computation of coarse-level mobilities, which is the primary bottleneck in the pure VE model. Although the gas plume distance requires comparatively little computational effort, it served as a useful case study to demonstrate how data-driven models can be employed to provide improved initial guesses for non-linear solvers. Notably, the code blocks targeted for substitution are precisely those that are called millions of times during a simulation, making the choice of data-driven models with short inference times essential.

\revision{For our applications,} linear regression and spline interpolation models provided the fastest predictions after training. This motivated their selection as surrogates. Since these models are simple, considerable effort was dedicated to identifying a minimal set of representative features that nevertheless capture the essential behaviour of the predicted quantities. Based on these features, we trained surrogates to provide enhanced initial guesses for the gas plume distance solver, to replace the computation of coarse-level mobilities, and to substitute the evaluation of secondary variables within the coupling scheme of the hybrid model. Importantly, the surrogates for the gas plume distance and coupling fluxes were designed to enhance, rather than replace, the original computations, thereby preserving mass balance while still reducing computational effort.

In the presented test cases, \revision{using data-driven surrogates resulted in a reduction of computational cost that ranged from 18\% to 75.4\%}, while maintaining mass conservation and introducing only negligible errors in the solution fields. With these improvements, the enhanced hybrid models achieved runtimes clearly below those of a corresponding FD simulation, demonstrating its viability as a competitive alternative to traditional FD modelling.

Looking ahead, possible extensions of this work include further augmentation of the VE scheme by training surrogates for the fine-level reconstruction step. Additionally, more advanced surrogates infused with physical knowledge may allow for the complete replacement of the coupling flux computations rather than mere enhancement. Data-driven models may also prove valuable in adaptive schemes, e.g. by predicting adaptivity criteria. Although the computation of adaptivity criteria, at least in 2D~\cite{buntic2025}, is less expensive than flux or mobility evaluation, their optimization through trained predictors could further improve performance.

In summary, this study demonstrates that data-driven methodology is a powerful tool for enhancing physical flow models. With careful substitution of selected components, it is possible to achieve substantial computational speedups while preserving essential physical properties such as mass conservation.

\backmatter

\bmhead{Acknowledgements}

This work was funded by Deutsche Forschungsgemeinschaft (DFG, German Research Foundation) under Germany's Excellence Strategy - EXC 2075 – 390740016. We acknowledge the support by the Stuttgart Center for Simulation Science (SimTech).

\section*{Declarations}






\bmhead{Code availability}
The code for reproducing the presented results can be accessed via \url{https://git.iws.uni-stuttgart.de/dumux-pub/buntic2025a} or \url{https://doi.org/10.18419/DARUS-5343}.


\begin{appendices}
\section{\revision{Closed solution for gas plume distance $z_p$}}\label{ch_closed_form_zp}
\revision{Given the mass balance within a column, see \cref{eq_gasPlumeDist_massBalance,eq_gasPlumeDist_integral_extended}, a closed form for $z_p$ only exists under special conditions, while in general $z_p$ needs to be deduced numerically. To derive a closed form, we first simplify \cref{eq_gasPlumeDist_integral_extended} by merging variables:
\begin{align}
    A &:= \frac{\left(1-s_{w,r}-s_{n,r}\right)p_e^\lambda}{\left(1+s_{w,r}\right)\left(1-\lambda\right)(\varrho_w-\varrho_n)g_z} \\
    B &:= p_e + \left(\varrho_w-\varrho_n\right)g_z z_T \\
    C &:= \left(\varrho_w-\varrho_n\right)g_z \\
    D &:= (1+s_{w,r})^{-1}\left(z_B + \frac{p_e\left(1-s_{w,r}-s_{n,r}\right)}{\left(1-\lambda\right)\left(\varrho_w-\varrho_n\right)g_z} - s_{w,r}z_T + S_w\left(z_T-z_B\right)\right).
\end{align}
Substituting these, reduces \cref{eq_gasPlumeDist_integral_extended} to
\begin{equation} \label{eq_zp_simplified}
    z_p + A\left(B-Cz_p\right)^{1-\lambda} = D,
\end{equation}
which is equal to the polynomial equation of degree $\lambda$:
\begin{equation} \label{eq_final_zp_form}
    y^\lambda - \left( B - CD\right)y^{\lambda-1} - AC = 0, 
\end{equation}
with $y=B-C z_p$. For integer values of $\lambda\in\{0,1,2,3,4\}$, according to the Abel-Ruffini theorem \cite{abel1826}, explicit formulas expressed in radicals exist. For our test cases, we consistently deployed $\lambda=2$, for which the solution of \cref{eq_final_zp_form} can be expressed as
\begin{equation}
    z_p = \frac{B+CD \pm \sqrt{\left(B-CD\right)^2 + 4AC}}{2C},
\end{equation}
with the subtracting branch being the desired solution. However, for integer values of $\lambda$ larger than 4 or non-integer values in general, \cref{eq_zp_simplified} needs to be solved numerically thus motivating our data-driven enhancement of the gas plume distance. Also note, that \cref{eq_gasPlumeDist_integral_extended} is only valid for Brooks-Corey relationships and may vary for other approaches.}

\section{Motivation for using linear regression and spline interpolation models} \label{ch_inferencetime}
\revision{In this appendix, we would like to compare the computational efforts of common data-driven regressors in order to motivate our choice of the linear regression and spline interpolation method.} The computational effort of machine learning models can be divided into three main components. The first is the training phase, i.e., the time required to fit a model to a given dataset. Since training is performed once prior to a simulation and the models are not updated during runtime, this contribution is less decisive. The second component is the time needed to transform simulation data into a format compatible with the machine learning model. Such transformations may include normalization or the computation of additional features from existing quantities. As these operations are performed at every model call, their cost is significant and must be minimized by keeping both the number of input parameters and the number of transformations per parameter small. The third component is the evaluation of the trained model, referred to as the inference time, which measures the time required to compute a prediction given a prepared set of input features. Since the surrogate models are evaluated many times during a simulation, inference time is as critical as data transformation cost. In some cases, postprocessing of the predictions may also be necessary, but in the present work this contribution was negligible and is therefore ignored. While input data transformation depends only loosely on the choice of machine learning method, inference time is directly tied to the specific method used and thus serves as a decisive criterion for model selection. The discussion in this section focuses exclusively on computational effort, without reference to accuracy or prediction quality.

To analyse the computational performance of different machine learning models, we construct an artificial dataset using the function
\begin{equation} \label{eq_artificial_target}
    f(x_1,x_2)=\ln(x_1+4) + e^{-x_2},
\end{equation}
with input parameters $x_1, x_2 \in [0,5]$. Although simple, \cref{eq_artificial_target} contains non-linear contributions that are non-trivial to approximate in trained models. Each input range is discretized into 51 equidistant points, yielding the sets $X_1$ and $X_2$. The Cartesian product $X_1 \times X_2$ defines the training set, with target values computed directly from \cref{eq_artificial_target}.

For comparison, we evaluate several widely used regression methods: Linear Regression (LR), Ridge Regression (RR), Decision Tree Regression (DTR), Random Forest Regression (RFR), Gradient Boosting Regression (GBR), Support Vector Machine Regression (SVMR), and a Multilayer Perceptron Regression (MLPR) model representing a neural network. All models were implemented using the default options in scikit-learn~\cite{scikit2011} and subsequently converted into ONNX format~\cite{onnx2021}, which provides faster inference due to its C++ implementation. For LR and RR, predictions can additionally be evaluated explicitly by directly multiplying the trained coefficients with the input data, bypassing both scikit-learn and ONNX. This manual implementation yields significantly faster evaluation. Henceforth, when referring to the linear regression model, we distinguish between the explicitly evaluated version, denoted as LR\_manual, and the backend-based versions. For the remaining models, explicit evaluation is not straightforward, so we rely on the respective interfaces and data structures provided by the libraries.

The required computational effort is summarized in \cref{table_inferences}. All models were trained on the same dummy dataset, consisting of two input parameters and one scalar target value. As an exemplary transformation, each input was mapped through the composition $h \circ g(x)=h(g(x))$, with $g(x)=e^x$ and $h(x)=\ln(x)$. This transformation effectively reproduces the original input values while introducing additional computational cost. In this simplified analysis, the transformation cost is identical across all methods and is therefore reported as the same elapsed time in \cref{table_inferences}. In practice, however, different data-driven models may benefit from different feature sets, leading to varying numbers of required transformations. By applying transformations in this analysis, we aim to mimic such a potential necessity. Each method was tasked with transforming and predicting $10^6$ values using the same set of input features. The corresponding computational effort is reported as the inference time in \cref{table_inferences}.
\begin{table}[h!]
    \centering
    \begin{tabular}{||c | c | c | c||}
        \hline
        Model & Training time [s] & Transform time [s] & Inference time [s] \\ 
        \hline\hline
        LR\_manual & 0.0006 & 0.2997 & 0.5933 \\ 
        \hline
        RR\_manual & 0.0007 & 0.2997 & 0.59664 \\ 
        \hline
        Spline & 0.0002 & 0.2997 & 2.4462 \\ 
        \hline
        LR & 0.0006 & 0.2997 & 6.8210 \\ 
        \hline
        RR & 0.0007 & 0.2997 & 6.8720 \\ 
        \hline
        DTR & 0.0032 & 0.2997 & 6.6107 \\ 
        \hline
        RFR & 0.2341 & 0.2997 & 12.1796 \\ 
        \hline
        GBR & 0.0736 & 0.2997 & 10.3144 \\ 
        \hline
        MLPR & 0.1568 & 0.2997 & 8.2414 \\ 
        \hline
        SVMR & 0.0048 & 0.2997 & 7.9865 \\ 
        \hline
    \end{tabular}
    \caption{Comparison of training time, input data transformation time and inference time for different, standard machine learning approaches and a 2D spline method.}
    \label{table_inferences}
\end{table}

The first observation is that training time is negligible in this context, as training is performed once prior to deployment and scales only with dataset size. Second, the transformation step requires less time than inference for all methods considered. However, if a larger number of transformations were needed, the transformation cost could become comparable to inference time, particularly for models with inherently short evaluations. As discussed earlier, inference time is strongly method-dependent, whereas transformation cost is less sensitive to the chosen model. Consequently, inference time will serve as the primary criterion for assessing the suitability of a data-driven approach. The results show that the random forest regressor is the slowest by a significant margin, followed by the gradient boosting and support vector machine regressors. In contrast, linear and ridge regression models exhibit the shortest inference times when evaluated manually, see the entries for LR\_manual and RR\_manual, as their trained coefficients can be directly applied to the inputs, bypassing library overhead. This explains their pronounced speedup relative to the embedded implementations. Among the tested methods, linear regression emerges as the fastest option and is therefore adopted for our simulation studies. It should be emphasized, that the performance comparison in \cref{table_inferences} applies only to the presented setup and should be regarded as a guideline rather than a universal ranking. The relative performance of the models may vary depending on the specific equations, feature sets, and problem scales.

In practical simulations, the number of inference calls depends on grid resolution and simulation end time, ranging from a few hundred to several hundred million calls. Given this scale, even small savings in per-inference cost translate into substantial overall gains. For this reason, we primarily employ linear regression whenever classical machine learning methods suffice. In cases where regression does not provide satisfactory accuracy, we fall back to 2D spline interpolation. Although slightly slower than manual linear regression, the spline approach remains considerably faster than most of the other tested methods. However, spline interpolation becomes impractical when more than two input features are required, as both training and prediction costs increase significantly with dimensionality.
\end{appendices}


\bibliography{sn-bibliography}

\end{document}